\documentclass[aps,preprint,noshowpacs,superscriptaddress,groupedaddress]{revtex4}
\usepackage{graphicx}
\usepackage[fleqn]{amsmath}
\usepackage{amssymb}
\usepackage[dvipsnames]{xcolor}
\usepackage[normalem]{ulem}
\definecolor{goodgreen}{rgb}{0.1,0.5,0}
\definecolor{goodred}{rgb}{0.7,0,0}
\usepackage{float}
\usepackage{multirow}
\usepackage[colorlinks,urlcolor=goodgreen,citecolor=blue,linkcolor=goodred]{hyperref}
\usepackage{bm}

\allowdisplaybreaks
\newcommand{\rem}[1]{} 

\begin{document}
\title{Nonlinear nanomechanical resonators approaching the quantum ground state

}
\author{C. Samanta}
\affiliation{ICFO - Institut De Ciencies Fotoniques, The Barcelona Institute of Science and Technology, 08860 Castelldefels (Barcelona), Spain}
\author{S. L. De Bonis}
\affiliation{ICFO - Institut De Ciencies Fotoniques, The Barcelona Institute of Science and Technology, 08860 Castelldefels (Barcelona), Spain}
\author{C. B. M{\o}ller}
\affiliation{ICFO - Institut De Ciencies Fotoniques, The Barcelona Institute of Science and Technology, 08860 Castelldefels (Barcelona), Spain}
\author{R. Tormo-Queralt}
\affiliation{ICFO - Institut De Ciencies Fotoniques, The Barcelona Institute of Science and Technology, 08860 Castelldefels (Barcelona), Spain}
\author{W. Yang}
\affiliation{ICFO - Institut De Ciencies Fotoniques, The Barcelona Institute of Science and Technology, 08860 Castelldefels (Barcelona), Spain}
\author{C. Urgell}
\affiliation{ICFO - Institut De Ciencies Fotoniques, The Barcelona Institute of Science and Technology, 08860 Castelldefels (Barcelona), Spain}
\author{B. Stamenic}
\affiliation{UCSB Nanofabrication Facility, ECE Department, Santa Barbara, CA 93106, USA}
\author{B. Thibeault }
\affiliation{UCSB Nanofabrication Facility, ECE Department, Santa Barbara, CA 93106, USA}
\author{Y. Jin}
\affiliation{C2N, CNRS, Université Paris-Saclay, 91120 Palaiseau, France}
\author{D. A. Czaplewski}
\affiliation{Center for Nanoscale Materials, Argonne National Laboratory, Argonne, IL 60439, USA}
\author{F. Pistolesi}
\affiliation{Universit\'{e} de Bordeaux, CNRS, LOMA, UMR 5798, F-33400 Talence, France}
\author{A. Bachtold}
\affiliation{ICFO - Institut De Ciencies Fotoniques, The Barcelona Institute of Science and Technology, 08860 Castelldefels (Barcelona), Spain}



\begin{abstract}
\textbf{An open question in mechanics is whether mechanical resonators can be made nonlinear with vibrations approaching the quantum ground state. This requires engineering a mechanical nonlinearity far beyond what has been realized thus far.
Here we discovered a mechanism to boost the Duffing nonlinearity by coupling the vibrations of a nanotube resonator to single-electron tunneling and by operating the system in the ultrastrong coupling regime.
Remarkably, thermal vibrations become highly nonlinear when lowering the temperature. The average vibration amplitude at the lowest temperature is 13 times the zero-point motion, with approximately 42\% of the thermal energy stored in the anharmonic part of the potential.
Our work paves the way for realizing mechanical `Schr\"{o}dinger cat' states~\cite{Kirchmair2013}, mechanical qubits~\cite{Rips2013,Pistolesi2021}, and quantum simulators emulating the electron-phonon coupling~\cite{Bhattacharya2021}.
}
\end{abstract}

\maketitle

Mechanical resonators are perfect linear systems in experiments carried out in the quantum regime. Such devices enable quantum squeezing of the mechanical motion \cite{Pirkkalainen2015a,Lecocq2015,Wollman2015}, quantum backaction evading measurements \cite{Suh2014,OckeloenKorppi2016,Moeller2017}, and entanglement between mechanical resonators \cite{Kotler2021,Wollack2022}. Achieving nonlinear vibrations in resonators approaching the quantum ground state would offer novel prospects for the quantum control of their motion. These include the development of mechanical qubits~\cite{Rips2013,Pistolesi2021} and mechanical cat states~\cite{Kirchmair2013}.  However, creating the strong nonlinearities required for this endeavour has been out of reach in all the mechanical systems explored thus far.

Mechanical resonators become nonlinear at large vibration amplitude $x$, when the Duffing (or Kerr) nonlinear term of the restoring force is sizable. Specifically, the restoring force reads $F=-m\omega_\mathrm{m}^2x-\gamma x^3$ with $m$ the mechanical eigenmode mass, $\omega_\mathrm{m}$ the resonance frequency, and $\gamma$ the Duffing constant. The Duffing constant is usually small but can be engineered using a force field gradient or a two-level system \cite{Bachtold2022}. Although mechanical systems have been operated in large field gradients \cite{Gloppe2014,Unterreithmeier2009} and strongly coupled to two-level systems~\cite{Kettler2021,O'Connell2010,Satzinger2018,Chu2018,Arrangoiz-Arriola2019,LaHaye2009,Pirkkalainen2013,Ma2021}, it has not been possible to create sizeable Duffing nonlinearities for vibrations approaching the quantum ground state.

We demonstrate a new mechanism to boost the vibration nonlinearity by coupling the mechanical resonator to single-electron tunneling (SET) in a quantum dot (Figs.~1a,b). The latter is operated as a degenerate two-level system fluctuating between two states with $N$ and $N+1$ electrons. 
The capacitive coupling to the mechanical mode is described by the Hamiltonian $H=-\hbar g n x/x_{\rm zp}$,
where $g$ is the electromechanical coupling, $n=0, 1$ the additional electron number in the quantum dot, and $x_{\rm zp}$ is the vibration zero-point motion. In the adiabatic limit, when the electron fluctuation rate is faster than the bare mechanical  frequency ($\Gamma_\mathrm{e} > \omega_\mathrm{m}^\mathrm{o}$), the fluctuations result in the nonlinear restoring force given by
\begin{equation}
    F_{\rm eff}=-\left[{m\omega_\mathrm{m}^\mathrm{o}}^2-\frac{1}{4x_{\rm zp}^2}\frac{(\hbar g)^2}{k_{\rm B}T}\right]x-\frac{1}{48x_{\rm zp}^4}\frac{(\hbar g)^4}{(k_{\rm B}T)^3}x^3
    \label{eq:forceeffective}
\end{equation}
when the electronic two-level system is degenerate (Fig. 1b; Sup. Info. Sec. I.D).
A striking aspect of the nonlinearity is its temperature dependence, since the nonlinear Duffing constant significantly increases when reducing temperature. 
The vibration potential can even become purely quartic in displacement, since the linear part of the restoring force vanishes~\cite{Pistolesi2015} at low temperature when $2k_\mathrm{B}T=\hbar g^2/\omega_\mathrm{m}^\mathrm{o}$. 
This can be realized for mechanical systems not in their motional ground state ($k_\mathrm{B}T>\hbar \omega_\mathrm{m}^\mathrm{o}$) by operating the system in the ultrastrong coupling regime when $g >\sqrt{2} \omega_\mathrm{m}^\mathrm{o}$.
%
%
Equation~1 also indicates that the measurement of a large decrease of $\omega_\mathrm{m}$ at low temperature is a direct indication of a strong nonlinearity.
%
A large number of experiments have been carried out where mechanical vibrations are coupled to SET  \cite{Knobel2003,Lassagne2009,Steele2009,Okazaki2016,Khivrich2019,Urgell2020,Wen2020,Blien2020,vigneau2021ultrastrong}, but the decrease of $\omega_\mathrm{m}$ has always been modest. 

Carbon nanotube electromechanical resonators (Fig.~1a) are uniquely suited for demonstrating a strong vibration nonlinearity. 
Its ultra-low mass gives rise to a large coupling $g$, which is directly proportional to 
$x_\mathrm{zp}=\sqrt{\hbar/2\omega_\mathrm{m}^\mathrm{o}m}$. Moreover, high-quality quantum dots can be defined along the nanotube by two p-n tunnel junctions that are controlled by electrostatic means.
Figure~1c shows a conductance trace featuring regular peaks associated with SET through the system.

A large dip in $\omega_\mathrm{m}$ is observed when setting the system on a conductance peak (Figs.~2a,b) where the electronic two-level system is degenerate. This is consistent with the vibration potential becoming strongly anharmonic. The decrease of $\omega_\mathrm{m}$ is enhanced at lower temperatures (Fig.~2c), indicating that the high temperature harmonic potential smoothly evolves into an increasingly anharmonic potential.
These data are well reproduced by the universal function predicted for $\omega_\mathrm{m}$, which depends only on the ratio $\epsilon_\mathrm{P}/k_\mathrm{B}T$ (Sup. Info. Eq. 44); here $\epsilon_\mathrm{P}=2\hbar g^2/\omega_\mathrm{m}^\mathrm{o}$.
Similar decrease in $\omega_\mathrm{m}$ was observed in two other devices  (Sup. Info. Sec. II.H).

These measurements reveal that the system is deep in the ultrastrong coupling regime. The universal temperature dependence of $\omega_\mathrm{m}$ enables us to quantify $g$ with accuracy. The largest coupling obtained from measurements at different conductance peaks is $g/2\pi=500\pm36$~MHz (black dots in Fig.~3a), corresponding to $g/\omega_\mathrm{m}^\mathrm{o}=16.5\pm1.2$. The coupling is consistent with the estimation $g/2\pi=547\pm185$~MHz obtained from independent measurements where the coupling is determined from the capacitive force associated with one electron added onto the quantum dot (purple line in Fig. 3a). The large uncertainty of this second estimation is related to the measurement of the eigenmode mass 
that enters the zero-point motion expression. Figures~3a-c show that the device is operated in the ultrastrong coupling regime $g > \omega_\mathrm{m}^\mathrm{o}$ and the adiabatic limit $\Gamma_\mathrm{e}>\omega_\mathrm{m}^\mathrm{o}$, which are necessary conditions to realize strong vibration anharmonicity.

We now turn our attention to the driven nonlinear resonant response of the mechanical mode (Figs.~4a,b). The spectral peak is asymmetric for vibration amplitudes as low as $x\simeq 40\times x_\mathrm{zp}$. We do not observe the usual hysteresis in the nonlinear response when the driving frequency is swept back and forth. Moreover, the nonlinear resonator has a decreasing responsivity for an increasing drive (Fig.~4c). These data agree with a model that takes into account the strong nonlinearity and thermal fluctuations; see the red lines in Figs.~4a,c. The lack of hysteresis is explained by the low amplitude of driven vibrations compared to the thermal displacement amplitude, an unusual regime for driven  nonlinear response measurements~\cite{Bachtold2022}. The behaviour of the responsivity arises from the thermal fluctuations that modify the spectral response 
of the driven nonlinear resonator~\cite{Bachtold2022}.
By comparing the model to the whole measured set of spectra spanning the linear-nonlinear crossover (Figs.~4a,c), we determine the coupling $g/2\pi=646\pm 217$~MHz. 
We obtain $g/2\pi=757\pm197$~MHz from the quadratic dependence of the resonant frequency on the driven vibrational amplitude for Duffing resonators (Fig.~4b), which remains approximately valid in the presence of thermal fluctuations provided that the driven vibration amplitude is sufficiently small.
%
These two values of $g$ are consistent with the two first estimates.

The vibrations become strongly nonlinear at low temperature for vibrations approaching the quantum ground state.
Figure~4d shows the fraction of the thermal energy stored in the nonlinear part of the vibration potential 
$ {\cal U}_{\rm NL}=
[\langle U_{\rm eff}(x)\rangle -m\omega_{\rm m}^2\langle x^2 \rangle/2]/\langle U_{\rm eff}(x)\rangle
$
where $U_{\rm eff}(x)$ is the total effective vibration potential created by the coupling.
The fraction is directly estimated from the measured decrease of $\omega_\mathrm{m}$ using the theory predictions of the coupled system (Sup. Info. Sec. I.F). 
The effect of the nonlinearity on the vibrations becomes increasingly important as the temperature is decreased, since a larger fraction of the thermal energy is stored in the nonlinear part of the potential (Fig.~4d).
The fraction $ {\cal U}_{\rm NL}$ becomes approximately 42\% at the lowest measured temperature where the average amplitude of thermal vibrations is $x_\mathrm{th}\simeq 13\times x_\mathrm{zp}$. Having the nonlinear part of the restoring force comparable to its linear part is extreme for mechanical resonators, especially when considering that the thermal vibration amplitude is so close to the zero-point motion.

We have demonstrated a mechanism to create a strong mechanical nonlinearity by coupling a mechanical resonator and a two-level system in the ultrastrong coupling regime. 
Mechanical resonators endowed with a sizeable nonlinearity in the quantum regime enable numerous applications.
Novel qubits may be engineered where the information is stored in the mechanical vibrations; such mechanical qubits are expected to inherit the long coherence time of the mechanical vibrations and may be used for manipulating quantum information~\cite{Rips2013,Pistolesi2021}. Mechanical ‘Schrodinger cat’ states –- non-classical superpositions of mechanical coherent states –- can also be formed~\cite{Kirchmair2013} with enhanced quantum sensing capabilities in the detection of force and mass. In practice, such systems may be realized by coupling the mechanical vibrations to double-quantum dots~\cite{Khivrich2019}. This approach preserves both the strong mechanical nonlinearities measured in this work~\cite{Pistolesi2021} and high mechanical quality factors~\cite{Urgell2020,Gardner2011}. Coupling mechanical vibrations to yet more quantum dots in a linear array may realize an analogue quantum simulator of small-size quantum materials~\cite{Bhattacharya2021}. Such a simulator could explore the rich physics of strongly correlated systems where the electron-electron repulsion is competing with the electron-phonon interaction.

\vspace{2cm}

{\bf Acknowledgements:} We acknowledge support from ERC Advanced Grant No. 692876, Marie Sklodowska-Curie grant agreement No. 101023289, MICINN Grant No. RTI2018-097953-B-I00, AGAUR (Grant No. 2017SGR1664), the Quantera grant (PCI2022-132951), the Fondo Europeo de Desarrollo, the Spanish Ministry of Economy and Competitiveness through Quantum CCAA and CEX2019-000910-S [MCIN/AEI/10.13039/501100011033], MCIN with funding from European Union NextGenerationEU(PRTR-C17.I1), Fundacio Cellex, Fundacio Mir-Puig, Generalitat de Catalunya through CERCA. Work performed at the Center for Nanoscale Materials, a U.S. Department of Energy Office of Science User Facility, was supported by the U.S. DOE, Office of Basic Energy Sciences, under Contract No. DE-AC02-06CH11357. 

\vspace{0.5cm}

{\bf Author contributions:} CS, SLDB and DAC fabricated the devices with contributions from BS and BT. CS and SLDB carried out the measurements with support from CBM, RTQ, WY and CU. YB provided the amplifier used in the measurements. FP developped the theory. CS, SLDB, FP and AB analyzed the data. CS, CBM, DAC, FP and AB wrote the manuscript with inputs from the other authors. AB supervised the work.

\begin{figure}[tb!]
	\begin{center}
		\includegraphics[scale=1.5]{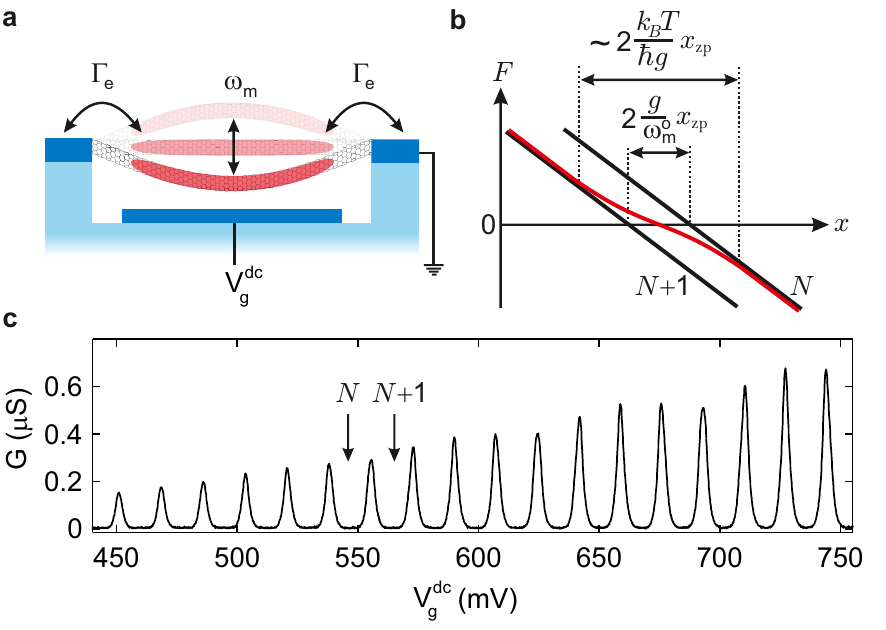}
		\caption{SET-based nonlinearity. (a) Schematics of the nanotube vibrating at $\omega_\mathrm{m}$. A quantum dot highlighted in red is formed along the suspended nanotube; the total electron tunnel rate to the two leads is $\Gamma_\mathrm{e}$. The vibrations are coupled to the electrons in the quantum dot via the capacitive coupling between the nanotube and the gate electrode. 
		(b) Origin of the SET-based nonlinearity. The two linear force-displacement curves shown in black correspond to the dot filled with either $N$ or $N+1$ electrons; the slope is given by the spring constant $m{\omega_\mathrm{m}^\mathrm{o}}^2$ and the two curves are separated by $\Delta x=2(g/\omega_\mathrm{m}^\mathrm{o})x_\mathrm{zp}$ caused by the force created by one electron tunnelling onto the quantum dot. 
		The force felt by the vibrations is an average of the two black forces weighted by the Fermi-Dirac distribution when $\Gamma_\mathrm{e} > \omega_\mathrm{m}^\mathrm{o}$.
		The resulting force in red is nonlinear for vibration displacements smaller than $\sim\frac{k_\mathrm{B}T}{\hbar g}x_\mathrm{zp}$; the reduced slope at zero vibration displacement indicates the decrease of $\omega_\mathrm{m}$. 
		(c) Gate voltage dependence of the conductance $G$ of device I at $T$=6~K. The average dot occupation increases by about one electron over the gate voltage range where a $G$ peak is observed.
		A voltage smaller than $k_\mathrm{B}T/e$ is applied to measure $G$.}
\label{fig1}
\end{center}
\end{figure}

\begin{figure}[tb!]
	\begin{center}
		\includegraphics[scale=1.5]{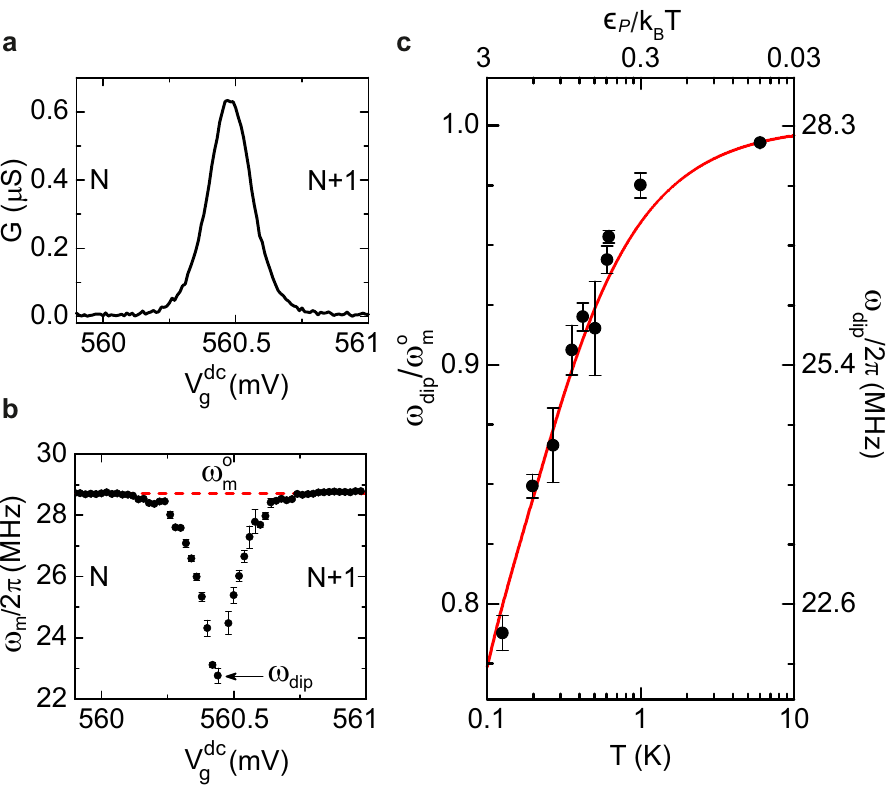}	
		\caption{Enhanced mechanical vibration nonlinearity at low temperature. (a,b) Conductance and mechanical resonance frequency as a function of gate voltage $V_\mathrm{g}^\mathrm{dc}$ at 300~mK. 
		By counting the number of observed conductance peaks from the nanotube energy gap, we estimate $N=22$. The red dashed line indicates $\omega_\mathrm{m}^\mathrm{o}$.  
		(c) Temperature dependence of the resonance frequency. The red solid line is the predicted universal function. The $\omega_\mathrm{dip}/\omega_\mathrm{m}^\mathrm{o}$ reduction is expected to be about $0.75$ when the potential is quartic; in this case, while the linear part of $F_{\rm eff}$ is zero, the nonlinear part of $F_{\rm eff}$ combined with thermal vibrations significantly renormalizes $\omega_\mathrm{m}$.  
		}
\label{fig2}
\end{center}
\end{figure}

\begin{figure}[tb!]
	\begin{center}
		\includegraphics[scale=1.2]{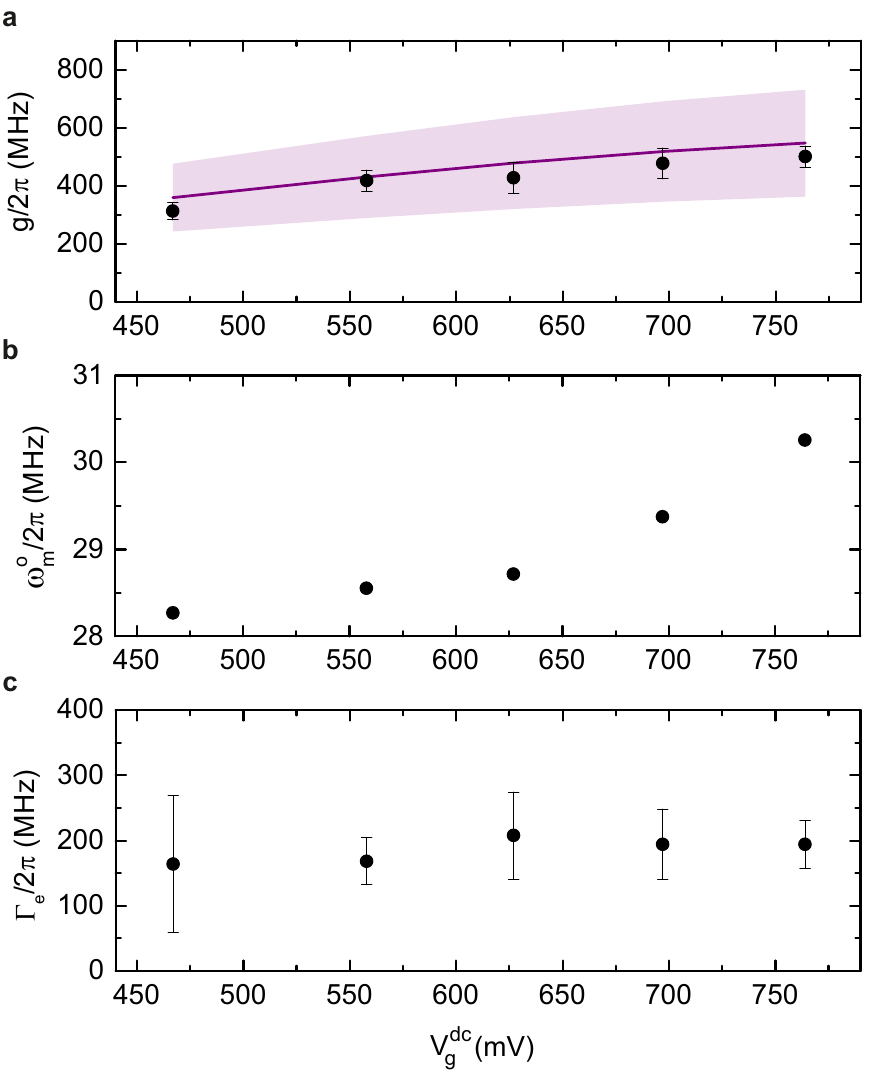}
		\caption{Electromechanical resonator in the ultrastrong coupling regime. (a) Coupling $g$ as a function of $V_\mathrm{g}^\mathrm{dc}$. The black data points are obtained from the measured temperature dependence of $\omega_\mathrm{m}$. The purple line is estimated from $g=e(C_\mathrm{g}'/C_\mathrm{\Sigma})V_\mathrm{g}^\mathrm{dc}/\sqrt{2m\hbar\omega_\mathrm{m}^\mathrm{o}}$ where $m$ is quantified from driven spectral response measurements 
		and where the spatial derivative of the dot-gate capacitance $C_\mathrm{g}'$ and the $V_\mathrm{g}^\mathrm{dc}$-dependent total capacitance $C_\mathrm{\Sigma}$ of the quantum dot are obtained from electron transport measurements. The uncertainty in this second estimation of $g$ is shown by the purple shaded area. 
		(b,c) Bare mechanical resonance frequency $\omega_\mathrm{m}^\mathrm{o}$ and electron tunnel rate $\Gamma_\mathrm{e}$ as a function of $V_\mathrm{g}^\mathrm{dc}$. The $\omega_\mathrm{m}^\mathrm{o}$ variation is due to the electrostatically-induced stress in the nanotube. We quantify $\Gamma_\mathrm{e}$ from the temperature dependence of the resonance width $\Delta\omega$ 
		in the spectral response measurements.
		%
		}
\label{fig3}
\end{center}
\end{figure}

\begin{figure}[tb!]
	\begin{center}
	\includegraphics[scale=1.2]{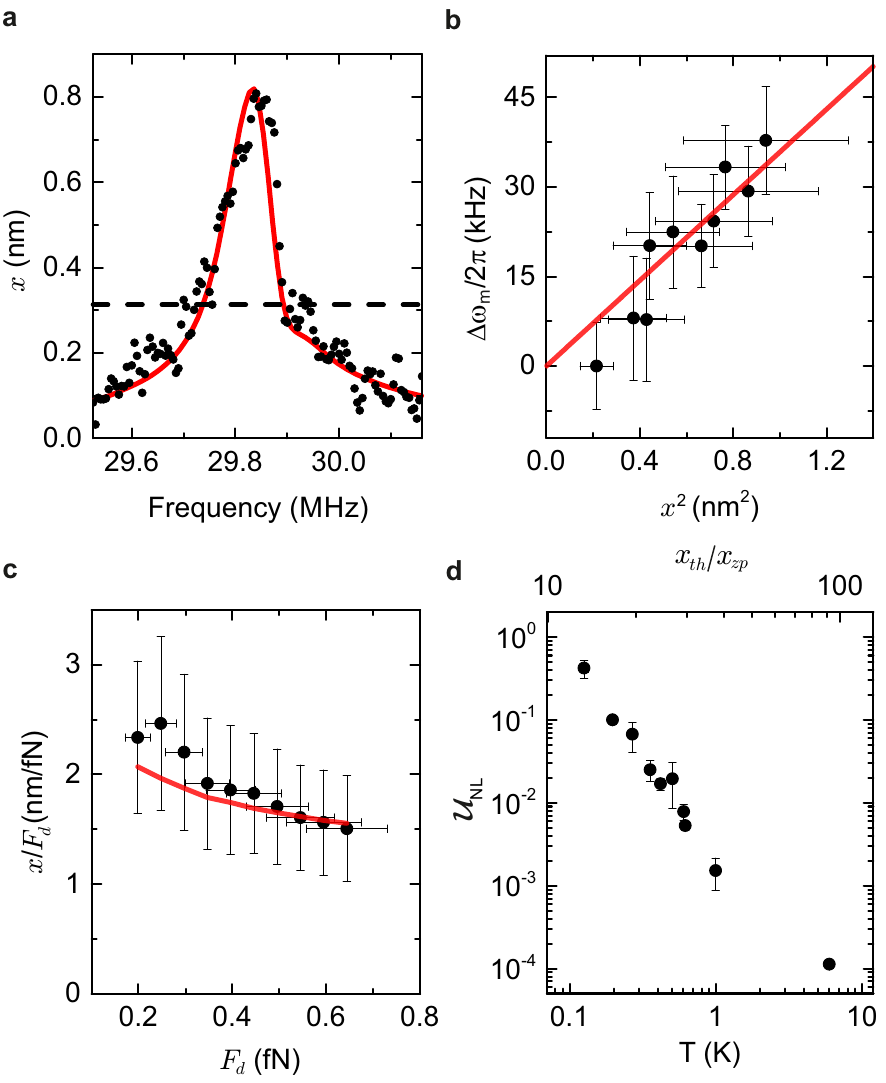}
	\caption{Nonlinear mechanical vibrations at $V_\mathrm{g}^\mathrm{dc}=757.2$~mV. (a) Nonlinear resonant response to the driving force at $6$~K. The dashed black line corresponds to $x=40\times x_\mathrm{zp}$. The red line is the simultaneous fit of 10 spectra at different drives, which span the linear-nonlinear crossover, to the theoretical prediction (Sup. Info. Sec. I.H). The response gets difficult to measure at lower $T$, since the vibration dissipation is enhanced and the drive is set by the oscillating gate voltage that has to be a few times smaller than the $V_\mathrm{g}^\mathrm{dc}$ width of the $G$ peak to avoid electrical nonlinearities.
	(b) Resonant frequency shift versus the driven vibration amplitude at $6$~K. The red line is a linear fit to the data. The driven vibration amplitude is set smaller or comparable to the averaged amplitude of the thermal vibrations $x_\mathrm{th}$. (c) Responsivity $x/F_\mathrm{d}$ of the mechanical mode at $6$~K, where $F_\mathrm{d}$ is the driven force amplitude. The red line is the fit to the theoretical prediction (Sup. Info. Sec. I.H). (d) Ratio ${\cal U}_{\rm NL}$ between the thermal vibration energy stored in the nonlinear part of the potential and that in the total vibration potential.
			}
\label{fig4}
\end{center}
\end{figure}

\end{document}



\title{ {\huge Supplementary Information}
\\
Nonlinear nanomechanical resonators approaching the quantum ground state
}
\author{C. Samanta}
\affiliation{ICFO - Institut De Ciencies Fotoniques, The Barcelona Institute of Science and Technology, 08860 Castelldefels (Barcelona), Spain}
\author{S. L. De Bonis}
\affiliation{ICFO - Institut De Ciencies Fotoniques, The Barcelona Institute of Science and Technology, 08860 Castelldefels (Barcelona), Spain}
\author{C. B. M{\o}ller}
\affiliation{ICFO - Institut De Ciencies Fotoniques, The Barcelona Institute of Science and Technology, 08860 Castelldefels (Barcelona), Spain}
\author{R. Tormo Queralt}
\affiliation{ICFO - Institut De Ciencies Fotoniques, The Barcelona Institute of Science and Technology, 08860 Castelldefels (Barcelona), Spain}
\author{W. Yang}
\affiliation{ICFO - Institut De Ciencies Fotoniques, The Barcelona Institute of Science and Technology, 08860 Castelldefels (Barcelona), Spain}
\author{C. Urgell}
\affiliation{ICFO - Institut De Ciencies Fotoniques, The Barcelona Institute of Science and Technology, 08860 Castelldefels (Barcelona), Spain}
\author{B. Stamenic}
\affiliation{UCSB Nanofabrication Facility, ECE Department, Santa Barbara, CA 93106, USA}
\author{B. Thibeault }
\affiliation{UCSB Nanofabrication Facility, ECE Department, Santa Barbara, CA 93106, USA}
\author{Y. Jin}
\affiliation{C2N, CNRS, Université Paris-Saclay, Palaiseau, France}
\author{D. A. Czaplewski}
\affiliation{Center for Nanoscale Materials, Argonne National Laboratory, Argonne, IL, 60439, USA}
\author{F. Pistolesi}
\affiliation{Universit\'{e} de Bordeaux, CNRS, LOMA, UMR 5798, F-33400 Talence, France}
\author{A. Bachtold}
\affiliation{ICFO - Institut De Ciencies Fotoniques, The Barcelona Institute of Science and Technology, 08860 Castelldefels (Barcelona), Spain}
\date{\today}

\maketitle

%
%
\newcommand{\beq}{\begin{equation}}
\newcommand{\eeq}{\end{equation}}
\newcommand{\beqa}{\begin{eqnarray}}
\newcommand{\eeqa}{\end{eqnarray}}
\newcommand{\rem}[1]{}
\newcommand{\refe}[1]{Eq.~(\ref{#1})}
%
%

\tableofcontents

\section{Theoretical description}
 \label{theory}
 
 \subsection{Central results}
 \label{Mainresults}
\begin{figure}[h]
\begin{center}
 \includegraphics[width=8.5cm]{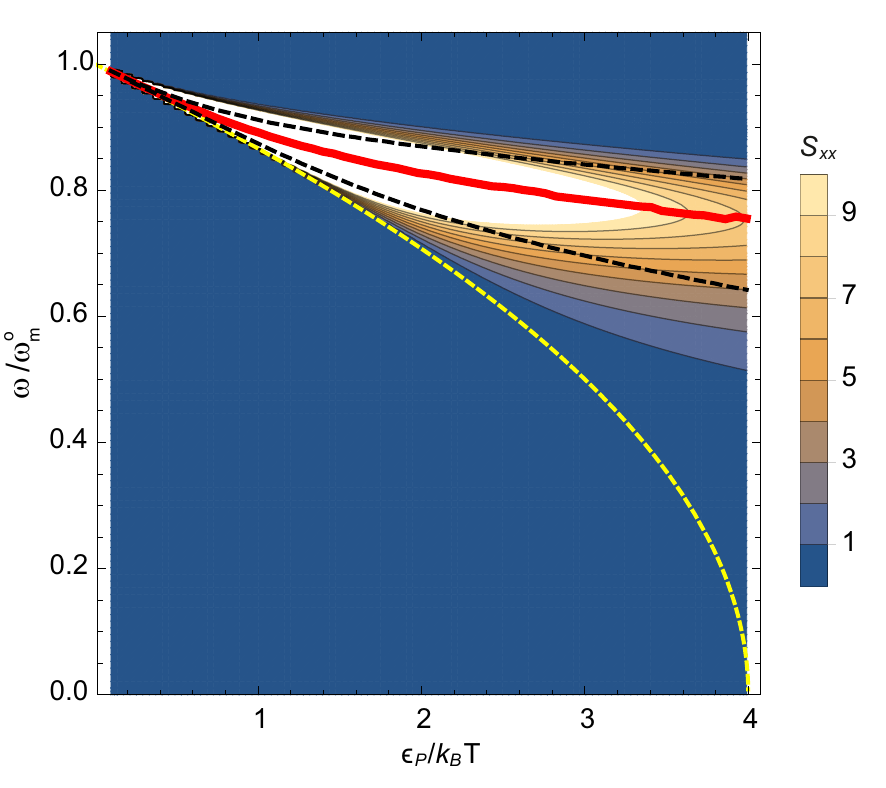}
 \end{center}
\caption{
Contour plot on the plane $\epsilon_P/k_B T$-$\omega$ of the intensity of the spectrum $S_{xx}(\omega)$ in units of $\epsilon_P/m\omega_0^3$.
The spectrum is evaluated at a conductance peak; this is the well-known peak arising from single-electron tunneling in quantum dots as the gate voltage is swept.
The thick continuous red line indicates the maximum of the spectrum, the dashed black lines indicates the values 
$\omega$ for which the intensity of the spectrum is reduced by a factor of 2 with respect to the maximum.
The yellow dotted line is the prediction of \refe{om4small} for the value of $\omega_Q$.
}
\label{FigSpectrum}
\end{figure}

We first highlight the main theoretical results that emerge from the coupling of a nanomechanical resonator coupled to a quantum dot operated in the incoherent single-electron tunneling regime. 
When the vibrations are slow with respect to the typical electronic tunneling rate, one finds that the effective force reads:
\beq
    F_{\rm eff}(x) = -m {\omega_m^o}^2 x + F_{\rm e} f_F(\epsilon-F_{\rm e} x)
    ,
\eeq
where $m$ is the eigenmode mass, $\omega_m^o$ is the bare resonance frequency, $F_e=\hbar g/x_\mathrm{zp}$ is the variation of the force acting on the mechanical system when the number of electrons in the dot varies by one unit, $\epsilon$ is the electron energy level, and $f_F$ is the Fermi-Dirac function.
One can define a resonance frequency from the quadratic term of the effective vibration potential obtained by the integration of $F_{\rm eff}$.
It reads
$\omega_Q=\omega_m^o (1-\epsilon_P/4 k_B T)^{1/2}$
where $\epsilon_P=2 \hbar g^2/\omega_m^o$ is the polaronic energy,
$T$ the temperature, and $k_B$ the Boltzmann constant.
Remarkably, the resonance frequency $\omega_Q$ associated with the linear restoring force decreases when lowering the temperature and vanishes at
$T=4\epsilon_P/k_B$.
The dependence of $\omega_Q$ as a function 
of $\epsilon_P/k_B T$ is shown as a dotted (yellow) line in Fig.~\ref{FigSpectrum}.

Another striking effect of the coupling and of the suppression of $\omega_Q$ is that the nonlinear part of the restoring force becomes dominant at low temperatures.
Due to this non-linearity, the period of oscillation becomes strongly dependent on the oscillation amplitude.
Thermal fluctuations allow the oscillator to explore different amplitudes, and thus different resonance frequencies: When averaged, these fluctuations lead to an observed resonance frequency that is much higher than $\omega_Q$ (see the red line in Fig.~\ref{FigSpectrum}).
In other words, the effect of the nonlinearity becomes more important when the vibrations are cooled to low temperature. This is just the opposite of what has been observed in mechanical resonators so far.

Despite the rich physics at work, the temperature dependence of the observed resonance frequency is a universal function of $\epsilon_P/k_B T$ for weak damping.
We find this by calculating the displacement fluctuation spectrum
$S_{xx}(\omega)$ (shown as a density plot in Fig.~\ref{FigSpectrum}).
It has been shown \cite{Micchi2015} that $S_{xx}(\omega)$ is proportional to the amplitude response to a weak drive, which is what we measure in this work. 
The temperature dependence of the measured resonance frequency in Fig.~2c of the main text agrees well with the prediction shown by the full red line in Fig.~\ref{FigSpectrum}. It is used to extract the value of $\epsilon_P$ and, therefore, $g$.
In the remaining part of the section~\ref{theory}, we use $\omega_0$ instead of $\omega_m^o$ for the bare resonance frequency to make the expressions more compact.

This section provides the basic theoretical description of the problem. 
It is structured as follows.
In Sec.~\ref{IncTrans} we will first recall the standard description of incoherent transport in a single-electron transistor.  
We then derive in Sec.~\ref{BackAction} the back-action of the electronic transport on the oscillator, obtaining an expression for the damping coefficient and the 
quadratic coefficient of the effective potential.
In Sec.~\ref{EffectivePotential} we derive in the equilibrium case the full non-linear effective potential of the oscillator.
We obtain $\omega_Q$ from the 
quadratic term of the effective potential.
In Sec.~\ref{SpectrumSoft} we derive the displacement fluctuation spectrum $S_{xx}(\omega)$ for the oscillator at equilibrium and discuss its behaviour. 
In Sec.~\ref{NonLinearCoef} we derive expressions for the first four coefficients in the expansion in the displacement $x$ of the effective potential and 
the average of the full non-linearity.
In Sec.~\ref{Duffing} we recall the main steps of the derivation of the Fokker-Planck equation for the oscillator quadratures in a Duffing description of the response to a drive. 
We then describe how we have used its numerical solution to fit the observed nonlinear spectra (Sec.\ref{Fitting procedure}).

 \subsection{Electronic transport}
 \label{IncTrans}
 
Electronic transport measurements in the device are described by incoherent tunnelling in the Coulomb blockade regime.
This corresponds to the condition $\hbar \Gamma_{\rm e} \ll k_B T \ll \Delta E \ll E_C$, where $\Gamma_{\rm e}$ is the tunneling rate of the electrons to the 
quantum dot, $E_C=e^2/C_\Sigma$ is the Coulomb energy ($C_\Sigma$ the total capacitance of the dot, $\Delta E$ the electronic level spacing, $e$ the electron charge).
For a description of transport in this regime see for instance Ref.~\cite{fazio_coulomb_2003}.
Concerning the oscillator we found that the system lays in the regime $\hbar \omega_0 \ll \hbar \Gamma_{\rm e} \ll k_B T$,
where we recall that $\omega_0$ is the (bare) mechanical resonance frequency measured far from the conductance peaks.
This implies that the oscillator can be described by a classical approach and that it is much slower than the electronic degrees of freedom.
We will use then a Born-Oppenheimer kind of approximation, where one first solves the electronic problem for a given value of the displacement 
$x$ of the oscillator and then considers the back-action of the electronic system on the oscillator
 \cite{armour_classical_2004,blanter_single-electron_2004,blanter_erratum_2005,mozyrsky_intermittent_2006,pistolesi_current_2007}.

 We begin with the electronic description for the case of $N_s$ electronic degrees of freedom,
 where $N_s=1$ describes the spinless case, 
 $N_s=2$ the spin- or valley-degenerate case, 
 and $N_s=4$ the case when both valley and spin degeneracy are taken into account. 
 We will find that most results do not depend on the actual value of $N_s$.
 In the incoherent transport regime 
 ($k_B T \gg \hbar \Gamma_{\rm e}$) the system is fully described by the probability that the electronic state 
 $\sigma$ (for instance $\sigma$ could be the spin projection) is occupied: $P_\sigma$. 
 We define also the probability that the dot is empty
 $P_0$, or occupied by one of the $\sigma$ states
 $P_1=\sum_\sigma P_\sigma$.
We assume that Coulomb blockade forbids double occupancy of the dot.
One can then write a Pauli master equation for these 
probabilities:
\beq
 	\dot P_\sigma=\Gamma^+ P_0 - \Gamma^- P_\sigma
 	.
\eeq
Here $\Gamma\pm=\Gamma_L^\pm+\Gamma_R^\pm$, and 
$\Gamma_\alpha^\pm$ is the rate at which an electron is added ($+$) or removed ($-$) from the quantum dot from the $\alpha$ (=$L$, $R$) lead.
The probability satisfies  the sum-rule $\sum_\sigma P_\sigma+P_0=1$.
The explicit expression of the rates depends on the Fermi distributions on the leads:
$\Gamma^+_\alpha=\Gamma_\alpha f_\alpha$ and $\Gamma^-_\alpha=\Gamma_\alpha (1-f_\alpha)$,
where by symmetry the rates do not depend on $\sigma$.
The electric current at the left lead and going from the left to the right lead, reads then:
\beq
	I_L=-e\left[N_s \Gamma_L^+(1-P_1)-\Gamma_L^- P_1\right]
\eeq
where we introduced the probability that the dot is occupied with one electron of any species $P_1=\sum_\sigma P_\sigma$. 
The equation of motion for the probability simplifies to 
\beq
	\dot P_1 = N_s \Gamma^+-(N_s \Gamma^++\Gamma^-) P_1
	.
\eeq
Using its stationary solution one finds that the current is 
\beq
	I_L=-e N_s  (\Gamma_R^- \Gamma_L^+-\Gamma_L^- \Gamma_R^+)/ \Gamma_{\rm e} ,
\eeq
where $\Gamma_{\rm e}=N_s\Gamma^+ + \Gamma^-$.
Setting the right lead to the ground as in the experiment and using $V_g$ and $V$ as the voltage potentials applied to the gate and the left lead, one can write the 
dependence of the Fermi distributions on these voltages: 
$f_L=f_F(\epsilon_1-e (V_g-V) C_g/C_\Sigma+e V C_R/C_\Sigma)$ and 
$f_R=f_F(\epsilon_1-e V_g C_g/C_\Sigma-e V C_R/C_\Sigma)$,  where
 $C_L$, $C_R$, $C_g$ are the left, right and gate capacitances,
$C_\Sigma=C_L+C_R+C_g$, $f_F(\epsilon)=1/(e^{\epsilon/k_B T}+1)$, and $\epsilon_1=\epsilon_0+(e^2-2Qe)/2C_\Sigma$, with $Q$ the total charge on the dot, and $\epsilon_0$ the single electron energy. 
 
 We can now calculate the conductance for vanishing bias voltage:
 \beq
 	G=-{\Gamma_L \Gamma_R N_s e^2 \over (\Gamma_L+\Gamma_R)[1+(N_s-1)f_0]}  {\partial f_0 \over \partial \epsilon_1}
 	\label{Gtunn}
\eeq
where we introduced $f_0=f_L=f_R$ for $V=0$.
The conductance has a maximum when the argument of the Fermi function equals $k_B T\ln N_s/2$.

\subsection{Back-action on the oscillator}
\label{BackAction}

 We now consider the coupling of the quantum dot to the oscillator. 
 When one electron is added to the dot, an additional electrostatic force $F_e$ acts on the 
 oscillator leading to a term in the Hamiltonian $-F_e x n$, where $n$ is the additional number of electrons on the dot. 
 In the incoherent regime $n$ can only take the integer values $0$ or $1$. 
 (In the main text we used the usual notation found in the opto-mechanical literature for the coupling $\hbar g = F_e x_{\rm zp}$,
 where $x_{\rm zp}=\sqrt{\hbar/2 m \omega_0}$ with $m$ the mass of the oscillator.)
 The presence of this coupling term changes the previous equations in Sec.~\ref{IncTrans} by introducing the following substitution in the rate expressions:
 \beq
 	\epsilon_0\rightarrow \epsilon_0-F_e x.
\eeq
We can then write a system of equations describing the oscillator position and the probability of occupation of the dot \cite{benyamini_real-space_2014}:
\beqa
	m \ddot x &=& -m \omega_0^2 x - m\gamma \dot x + F_e P_1(t) + F(t)
	\\
	\dot P_1(t)&=&N_s \Gamma^+(x)-\Gamma_{\rm e}(x) P_1
	,
\eeqa
here  $\gamma$ and $F(t)$ are the intrinsic damping rate and a weak external force driving the oscillator. 

Assuming small displacements from the equilibrium value of both $x$ and $P_1$ we can define 
\beqa
	x &=& x_0+ \tilde x(t) \\
	P_1 &=&P_1^{(0)}+\tilde P_1(t).
\eeqa
We obtain the following non-linear equations for $x_0$ and $P_1^{(0)}$:
\beqa
	x_0 &=& {F_e\over m \omega_0^2} P_1^{(0)}  \\
	P_1^{(0)} &=& {N_s \Gamma^+(x_0)\over \Gamma_{\rm e}(x_0) }, 
\eeqa
and a linear system for the small fluctuating parts $\tilde x$ and 
$\tilde P_1$:
\beqa
	 \ddot {\tilde x} &=&-\omega_0^2 {\tilde x} - \gamma  \dot {\tilde x} +(F_e/m) {\tilde P}_1 + F(t)/m\\
	{\tilde P}_1 &=& [N_s \partial_x \Gamma^+-P_1^{(0)} \partial_x \Gamma_{\rm e} ] {\tilde x} - \Gamma_{\rm e} {\tilde P}_1
	.
\eeqa
Introducing the Fourier transform ${\tilde x}(\omega)=\int dt e^{i \omega t} {\tilde x}(t)$ and using the explicit form of $P_1^{(0)}$ we have:
\beq
	\tilde P_1(\omega)(\Gamma_{\rm e}-i\omega)=N_s(\Gamma^-\partial_x \Gamma^+ -\Gamma^+\partial_x \Gamma^- ) {\tilde x}(\omega)
	.
\eeq
Substituting this expression into the equation for the displacement we have
\beq
	[-\omega^2+\omega_Q^2-i\omega \gamma_R ]{\tilde x}(\omega)=F(\omega)/m
\eeq
with the renormalized damping and frequency:
\beqa
	\gamma_R&=&\gamma-{F_e^2 N_s \over m \Gamma_{\rm e}} \Xi ,
	 \\
	\omega_Q^2 &=&\omega_0^2 +	
	{F_e^2 N_s \over m }  \Xi
	,
\eeqa
where we use the notation $\omega_Q$ since it is related to the quadratic coefficient of the 
effective potential. 
We defined
\beq
	\Xi= {\Gamma^-\partial_{\epsilon_1} \Gamma^+ -\Gamma^+\partial_{\epsilon_1} \Gamma^-\over \omega^2+\Gamma_{\rm e}^2} .
\eeq
More explicitly, we have:
\beq
	\Xi  =-{1\over k_B T}  {(\Gamma_L+\Gamma_R)[\Gamma_L f_L (1-f_L)+\Gamma_R f_R(1-f_R)] \over \omega_0^2 + \Gamma_{\rm e}^2}
	,
\eeq
where we substituted $\omega$ by the value of the unperturbed resonance frequency $\omega_0$.
Strictly speaking when the reduction of $\omega_Q$ is large one should insert $\omega_Q$ and obtain a self-consistent equation, but since we are interested in the limit of $\Gamma\gg \omega_0>\omega_Q$ this will have a negligible impact on the final result.
 The expression simplifies further in the relevant limit $eV \ll k_B T$ used in the experiment:
 \beq
	\Xi  =-{1\over k_B T}  {(\Gamma_L+\Gamma_R)^2  f_0 (1-f_0) \over \omega_0^2 + (\Gamma_L+\Gamma_R)^2[1+(N_s-1) f_0]^2}
	,
\eeq
where $f_0=f_L=f_R$ is the Fermi distribution of the leads.
For the damping we thus obtain
\beq
	\gamma_R=\gamma+{ \epsilon_P \over k_B T } {N_s \omega_0^2 (\Gamma_L+\Gamma_R) f_0(1-f_0) 
	\over k \left(\omega_0^2 + (\Gamma_L+\Gamma_R)^2 k^2\right)}
	,
	\label{dampo}
\eeq
with $k=1+(N_s-1) f_0$. We introduced the polaronic energy $\epsilon_P=F_e^2/m\omega_0^2=2 \hbar g^2/\omega_0$, which is the crucial energy scale of the problem.
For $\omega_0\ll \Gamma_L+\Gamma_R$ the resonance frequency simplifies to
\beq
	\omega_Q^2=\omega_0^2 \left[1-{\epsilon_P \over k_B T} N_s { f_0(1-f_0) \over [1+(N_s-1)f_0]^2} \right]
	.
\eeq
The importance of the ratio $\epsilon_P/k_B T$ it is clearly visible here, since when this ratio is sufficiently large $\omega_Q^2$ changes sign, indicating the appearance of a bistability.
Note that $f_0$ is the Fermi distribution of the metal electrodes. 
One can readily verify that the resonance frequency associated with the linear part of the restoring force is lowest when 
$f_0=1/(1+N_s)$. It reads
\beq
	\left. \omega_Q^2 \right|_{\rm min}=\omega_0^2 \left[1-{\epsilon_P \over 4 k_B T}\right]
	,
\eeq
independently of $N_s$.
Substituting the same value of $f_0=1/(N_s+1) $
into the expression for the 
damping \refe{dampo} for $\omega_0\ll \Gamma_L+\Gamma_R$ one obtains
\beq
\label{SETdissipation}
	\gamma_R=\gamma+{ \epsilon_P \over k_B T } 
	 {\omega_0^2 
	\over 4 \Gamma_{\rm e}}
	,
\eeq
where $\Gamma_{\rm e}=(\Gamma_L+\Gamma_R) N_s/(N_s+1)$.
Thus in terms of $\Gamma_{\rm e}$ the expression of the damping at the 
maximum of reduction of the frequency does not depend on $N_s$.
This expression has been used to extract the value of 
$\Gamma_e$ in the main text.

\subsection{Effective potential}
\label{EffectivePotential}

The reduction of the resonance frequency is due to the back-action of the electronic system on the oscillator. 
This generates an effective force acting on the oscillator that depends on $x$ in a non-linear fashion due to the
$x$-dependence of the tunnelling rate:
\beq
	F_{\rm eff}(x) = - m \omega_0^2 x+F_e N_s \Gamma^+(x)/\Gamma_{\rm e}(x).
\eeq
Here $\Gamma^+(x)$ and $\Gamma_{\rm e}(x)$ are defined in Sec.~\ref{IncTrans} using $\epsilon_0\rightarrow \epsilon_0-F_e x$. In the equlilibrium case ($eV\ll k_B T$) the force reads:
\beq
	F_{\rm eff}(x) = - m \omega_0^2 x+  { F_e \over e^{(\epsilon-F_e x)/k_B T-\ln N_s} +1 } ,
	\label{Feff}
\eeq
where $\epsilon=\epsilon_1-eV_g C_g/C_\Sigma$.
The electronic contribution is clearly non-linear.
The interpretation is simple. 
The force induced by the electrons is just $F_e$ multiplied by the probability that the dot is occupied by an additional 
electron. 
In equilibrium this probability is given by the Fermi function.
Note however that it does not coincide with the Fermi distribution of the metals ($f_0$), since the chemical potentials in the dot and in the leads differ. 
The number of electronic degrees of freedoms $N_s$ appear only 
as a shift of the energy level.
The equilibrium condition for the mechanical oscillator $F_{\rm eff}(x_0)=0$ gives 
\beq
	m \omega_0^2 x_0 = { F_e \over e^{(\epsilon-F_e x_0)/k_B T-\ln N_s}+1}
	.
\eeq
The spring constant is proportional to the derivative with respect to $x_0$ of the right-hand side of this expression.
It is maximal for 
\beq
	(\epsilon_M-F_e x_0)/k_B T=\ln N_s.
	\label{epsilonM}
\eeq 
When sweeping the gate voltage, that is $\epsilon$, the resonance frequency reaches it minimum at $\epsilon_M$. 
(One can show that the energies $\epsilon$ corresponding to the maximum of the conductance and to the the maximum of the reduction of the frequency coincide only in the case of 
spin-less fermions $N_s=1$; the difference in $\epsilon$ is however of the order of $k_B T \ln N_s$ and is thus difficult to resolve experimentally.)
In terms of the displacement from the equilibrium value, $\tilde x=x-x_0$, the effective force acquires a particularly simple form:
\beq
	F_{\rm eff}(\tilde x) =-m \omega_0^2 \tilde x +{F_e\over 2} \tanh\left[ {F_e \tilde x \over 2 k_B T}\right]
	\label{Feff}
	.
\eeq
We can then obtain the effective potential by integration:
\beq
	U_{\rm eff}(\tilde x) = {m \omega_0^2 {\tilde x}^2 \over 2}
		-{k_B T} \ln \left[ \cosh\left[ {F_e \tilde x \over 2 k_B T}\right]\right]
		\label{UeffSym}
		,
\eeq
where we choose the arbitrary potential constant such that 
$U_{\rm eff}(0)=0$.
The potential is symmetric in this case  [when \refe{epsilonM} holds], the general form can be readily derived by integrating  \refe{Feff}. 
%


From \refe{UeffSym} one can see that the effective potential evolves from a purely parabolic behaviour for $F_e$ small to a double well for $F_e$ sufficiently large. 
It is interesting to write the potential in terms of the dimensionless variable $z=\tilde x/(F_e/m\omega_0^2)$, that 
measures the distances in units of the displacement induced by the force $F_e$.
The potential reads:
\beq
	{U_{\rm eff} \over \epsilon_P}
	={z^2 \over2} -{1\over \tilde\epsilon_P} \ln \cosh(\tilde \epsilon_P z/2)
	\label{Ueff2}
	.
\eeq
\begin{figure}[h]
\begin{center}
 \includegraphics[width=8cm]{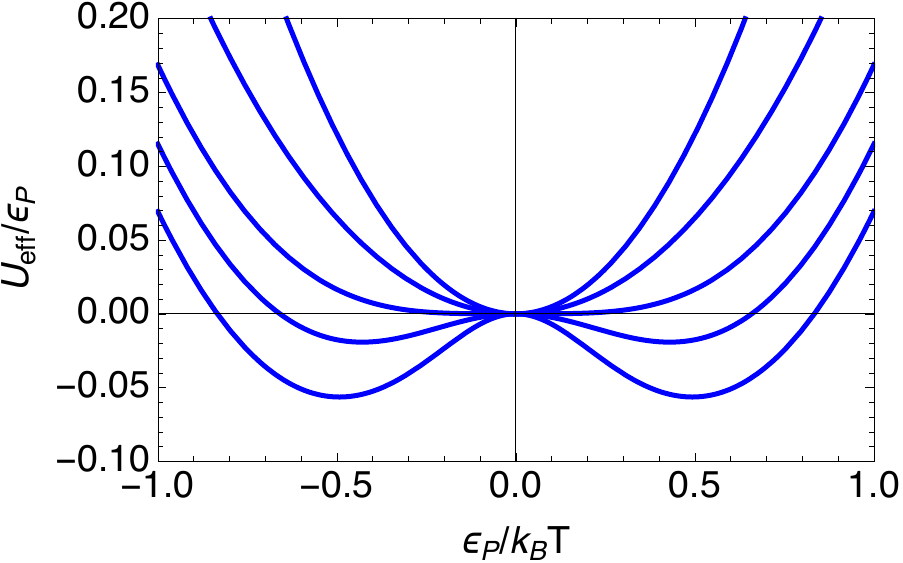}
 \end{center}
\caption{
Evolution of the effective potential for the oscillator for different values of the 
ratio $\epsilon_P/k_B T=0.1, 2, 4, 6, 10$ 
from the upper to the lower curve. 
For $\epsilon_P/k_B T=4$ the potential is quartic at leading order.
}
\label{UeffFig}
\end{figure}
One can see that its form depends now on a single parameter $\tilde \epsilon_P=\epsilon_P/k_B T$, 
that is the natural coupling constant of the problem. 
We show in Fig.~\ref{UeffFig} the evolution of the potential for $\tilde \epsilon_P=0.1, 2, 4, 6, 10$.
One can expand the potential to order 4  to obtain:
\beq
\label{Ueffexpansion}
	{U_{\rm eff} \over \epsilon_P}
	={z^2 \over2}\left(1-{\tilde\epsilon_P\over 4}\right) 
	+ {{\tilde \epsilon_P}^3 \over 192} z^4 + \dots \, .
\eeq
For $\tilde \epsilon_P \ll 1$, one has a simple harmonic oscillator. 
For $\tilde\epsilon_P=4$ the quadratic term vanishes and for small displacement the potential is quartic at leading order. 
For $\tilde\epsilon_P >4$ the system is bistable and features a double-well potential. 

The bistability is directly related to the two possible states of the dot, empty or filled with one electron. 
The phase diagram and the crossover to the bistability in the coherent tunnelling limit has been discussed in Refs.~\cite{micchi_mechanical_2015,micchi_electromechanical_2016}.
In the bistable region the current is strongly reduced leading to a current blockade induced by the electron-phonon coupling.
Recently it has been proposed to exploit this kind of bistability 
in a double-dot coupled to an oscillator to design a nanomechanical qubit\cite{pistolesi_proposal_2021}.

\subsection{Fluctuation spectrum and softening of the mechanical mode}
\label{SpectrumSoft}

The first effect of the coupling is a reduction of the resonance frequency. 
For small $\tilde\epsilon_P$ or for small displacement this follows from the expression of the quadratic part of the effective potential (Eq.~\ref{Ueffexpansion}) that leads to 
\beq
	{\omega_Q^2 \over \omega_0^2} =1-{\tilde\epsilon_P\over 4}
	\qquad
	{\rm for\ } \tilde\epsilon_P \ll1 
	\label{om4small}
	.
\eeq
This effect has been observed by several groups
\cite{Lassagne2009coupling,Steele2009,benyamini2014real,Khivrich2019,Urgell2020,Wen2020,Blien2020}.
For larger values of $\tilde\epsilon_P$ one cannot rely anymore on just the value of the quadratic part to quantify the observed 
mechanical resonance frequency.
The oscillator becomes highly non-linear, so some care has to be taken to measure the resonance frequency of the system in equilibrium.
This can be done by measuring the driven spectrum by keeping the driven vibration amplitude smaller than the standard deviation of the thermal vibration amplitude. 
Otherwise, the resonance frequency depends on the intensity of the drive used to detect it, see Sec.~\ref{Duffing}. 
Even for infinitesimal drive, the thermal fluctuations allow to explore regions of different vibration amplitudes for which 
the period of the oscillator takes values that can be very different. 
In order to find an averaged value for the observed resonance frequency for 
small drive we will follow again Ref.~\cite{micchi_mechanical_2015} and
calculate the displacement fluctuation spectrum at equilibrium:
\beq
	S_{xx}(\omega) = \int dt e^{i\omega t} \langle (\tilde x(t)-\langle\tilde x\rangle) (\tilde x(0)-\langle\tilde x\rangle)\rangle
	.
\eeq
For a small coupling constant $S_{xx}(\omega)$ reduces to a Lorentzian function peaked at $\omega_Q$ as defined 
in \refe{om4small}. 
For a larger coupling constant the resanance peak broadens and shifts to lower frequencies, 
but it remains well identified, and the resonance frequency can be determined for instance, by measuring 
$S_{xx}(\omega)$ \cite{moser_ultrasensitive_2013}.
In Ref.~\cite{micchi_mechanical_2015} it is shown that  in the equilibrium limit $S_{xx}(\omega)$ coincides with the response function 
for a weak drive, which is what is measured in this work. 
In equilibrium and for infinitesimal damping, $S_{xx}(\omega)$ can be calculated following the methods of statistical mechanics \cite{dykman_fluctuations_1980}:
\beq
	S_{xx}(t)=\int d\tilde x_o dp_o P(\tilde x_o,p_o) \left[\tilde x(t) \tilde x(0)-\langle \tilde x\rangle^2\right]
,
\eeq
where $\tilde x(t)$ is the solution to the time evolution of the displacement when the force is given by 
$F_{\rm eff}$ in \refe{Feff} with initial conditions for the displacement and the momentum $\tilde x_o$ and $p_o$.
The quantity $P$ is the Boltzman distribution:
\beq
	P(\tilde x_o,p_o)=  {\cal N} e^{-{p_o^2/2m+U_{\rm eff}(\tilde x_o) \over  k_B T}}
\eeq
where $U_{\rm eff}$ is given by \refe{UeffSym} and $\cal N$ is a normalization factor.

In order to perform the calculation it is convenient to change the integration variables. Instead of using $(\tilde x_o, p_o)$ we will use 
$(E,\tau)$, where $E=p_o^2/2m + U(\tilde x_o)$ and $\tau$ is the time along the trajectory of energy $E$.
The Jacobian is unitary $d\tilde x_o d p_o = dE d\tau$.
We can now write the spectrum as follows:
\beq
	S_{xx}(t) = \int dE  \int_0^{T_E} d\tau_E  e^{-E/k_B T} {\cal N} \tilde x_E(t+\tau_E) \tilde x_E(\tau_E) 
	\label{SxxE}
\eeq
where $\tau_E$ is the time along the trajectory with energy $E$ and $T_E$ is the period of the orbit.
Note that one could have more than one orbit for a given energy. We will focus in the following 
on the stable case occurring when $\tilde \epsilon_P<4$ and for which only one orbit is present.
%
We can now perform the Fourier transform of \refe{SxxE} by introducing the Fourier series of the orbit displacement:
\beqa
	\tilde x_{E}(\tau) &=& \sum_{n=-\infty}^{+\infty} e^{-i n \omega_{E} \tau} \tilde x_n(E)
	\\
	\tilde x_n(E) & =&  \int_0^{T_{E}} {d\tau\over T_{E}}  e^{i n \omega_{E}\tau  } x_{E}(\tau)
\eeqa
with $\omega_{E}=2\pi/T_{E}$.
Substituting these expressions into \refe{SxxE} and performing the Fourier transform we obtain:
\beq
	S_{xx}(\omega) =  {\cal N} \int dE    e^{-E/k_B T}  
	\sum_{n\neq 0} \left|  \tilde x_{n}(E)\right|^2  T_{E} 2 \pi \delta(\omega-n\omega_{E}) 
	\label{SxxE2}
	.
\eeq
Note that dropping the $n=0$ harmonics allows to subtract the average of the trajectory. 

We now perform the integral in the energy variable. The equation
$
	n \omega_{E_n(\omega)}=\omega
$
defines a function $E_{n}(\omega)$ for each trajectory.
We can then write:
\beq
	S_{xx}(\omega) =  {\cal N}    e^{-E_{n}(\omega)/k_B T} 
	\sum_{n\neq 0} \left|  \tilde x_{n}(E_{n}(\omega))\right|^2  {2 \pi n \over \omega}  {2 \pi \over n\left| {d \omega_{E}\over dE} \right|}  
	\label{SxxE3}
\eeq
This expression can be used to compute the spectrum  either numerically for any value of the parameters, or analytically in some limits.
In Fig.~\ref{FigSpectrum} we show the result of the numerical evalution of this expression for $\tilde \epsilon_P<4$. 
For $\tilde \epsilon_P>4$ the system becomes bistable; the crossover between the stable and the bistable regions occurs when the quadratic term of the effective potential vanishes, see the dotted yellow line in Fig.~\ref{FigSpectrum}.
Due to the strong non-linearity of the potential combined with the thermal fluctuations, the spectrum has a maximum corresponding to the resonance frequency (thick red continuous line), which approaches $\approx 0.75 \omega_0$ at $\tilde \epsilon_P=4$. 
The spectrum in Fig.~\ref{FigSpectrum} also shows a large broadening of the resonance due to phase fluctuations. The effective quality factor approaches 
$Q\simeq 5.5$ at $\tilde \epsilon_P=4$.
Note that there is no direct contribution of the dissipation to the peak broadening in the model. 
Taking into account the dissipation induced by the coupling between vibrations and single-electron tunneling (SET) changes only qualitatively the peak broadening in Fig.~\ref{FigSpectrum}.

The dependence  of the maximum of this spectrum as a function of  $\tilde \epsilon_P$ has been used in the main text (see Fig. 2c) to fit the 
temperature dependence of the resonance frequency at the gate voltage corresponding to the conductance peak and infer an estimate of $\epsilon_P$.
We do not have an analytical expression, but fitting the numerical result one obtains
\beq
    \omega_M/\omega_0=1+\sum_{n=1}^5 a_n {\tilde\epsilon_P}^n
\label{eq:relationfrequencyepsilon}
\eeq
with
$a_1=-0.127655$, 
$a_2=0.010475$, 
$a_3=0.0125029$, 
$a_4=-0.00480876$, 
and
$a_5=0.000515142$,
which is within 0.1\% of the numerical result for $0 \leq{\tilde\epsilon_P} \leq 4$.

\subsection{Coefficients of a series expansion of the potential in the displacement and estimation of the thermal energy stored in the non-harmonic part of the potential} 
\label{NonLinearCoef}

In this section we derive explicit expressions for the first 4 coefficients of the 
series expansion of the effective potential for small $\tilde x$. 
We  will express these nonlinear coefficients as well as the amount of thermal energy stored in the nonlinearity as a function of $\epsilon_P$. 
This allows us to quantify the amount of thermal energy stored in the nonlinearity shown in Fig.~4d from the suppression of the resonance frequency measured at each temperature in the main text using Eq.~\ref{eq:relationfrequencyepsilon}.   
We will use the standard notation:
\beq
	U_{\rm eff}( \tilde x) = U_0 
	+ {m \omega_Q^2 \over 2} {\tilde x}^2
	+ {m \beta_D \over 3} {\tilde x}^3
	+ {m \gamma_D \over 4} {\tilde x}^4
	.
	\label{standardNL}
\eeq
Since the constant is irrelevant, we can obtain the other coefficients directly from the expression of the 
effective force \refe{Feff} using
$	{dU_{\rm eff}/ dx} = -F_e n + m \omega_0^2 x$, ${d^2U_{\rm eff}/dx^2} = -F_e {dn/dx} + m \omega_0^2$, and so on.
Here 
$ n = 1/\left( \exp\{(\epsilon-F_e x)/k_B T-\ln N_s\} +1\right)$
and 
has to be evaluated at $x=x_0$, that is, the equilibrium position.
Using the properties of $n$ we have:
\beqa
	{\omega_Q^2 \over \omega_0^2} &=&	1- {\epsilon_P \over k_B T} n(1-n) \\
	\beta_D & =&  {F_e^3 \over 2 m (k_B T)^2} n (1-n)(2n-1) \\
	\gamma_D &=& - {F_e^4 \over 6 m (k_B T)^3} n(1-n)(6n^2-6n+1)
	\label{gammaDuffingfirst}
	.
\eeqa
We evaluate explicitly these expressions at the symmetric point for which $n=1/2$:
\beqa
	{\omega_Q^2 \over \omega_0^2} &=&	1- {\epsilon_P \over 4 k_B T} ,\qquad\beta_D  =  0, \\
	\gamma_D  &=&  {F_e^4 \over 48 m (k_B T)^3} = {\epsilon_P^2m \omega_0^4  \over 48 (k_B T)^3}
	\label{gammaDuffing}
	.
\eeqa
These expressions are independent of the value of $N_s$.


In order to quantify the degree of non-linearity of the potential 
it is interesting to compare the contribution of the average of the quadratic term of the potential 
to the average of the full effective potential. 
For this we can use the expression of the effective potential given by \refe{Ueff2}. 
The average value of $z^2$ reads:
\beq
	\langle z^2\rangle 
	=
	\int dz e^{-U_{\rm eff}(z)/{\tilde \epsilon_P}}  z^2
	\bigg/
	\int dz e^{-U_{\rm eff}(z)/{\tilde \epsilon_P}} 
	=1/\tilde \epsilon_P+1/4
	.
\eeq
Thus the average of the (dimensionless) quadratic part of the potential reads:
\beq
	U_2 \equiv {d^2 U_{\rm eff}\over dz^2}  \left\langle {z^2\over 2} \right\rangle 
	=(1-\tilde \epsilon_P/4)(1/\tilde \epsilon_P+1/4)/2
	.
\eeq
In a similar way we can calculate the average of the full potential
\beq
    \langle U_{\rm eff} \rangle = \int dz e^{-U_{\rm eff}(z)/{\tilde \epsilon_P}}  U_{\rm eff}(z)
	\bigg/
	\int dz e^{-U_{\rm eff}(z)/{\tilde \epsilon_P}} 
	.
\eeq

\begin{figure}[h]
\begin{center}
 \includegraphics[width=8cm]{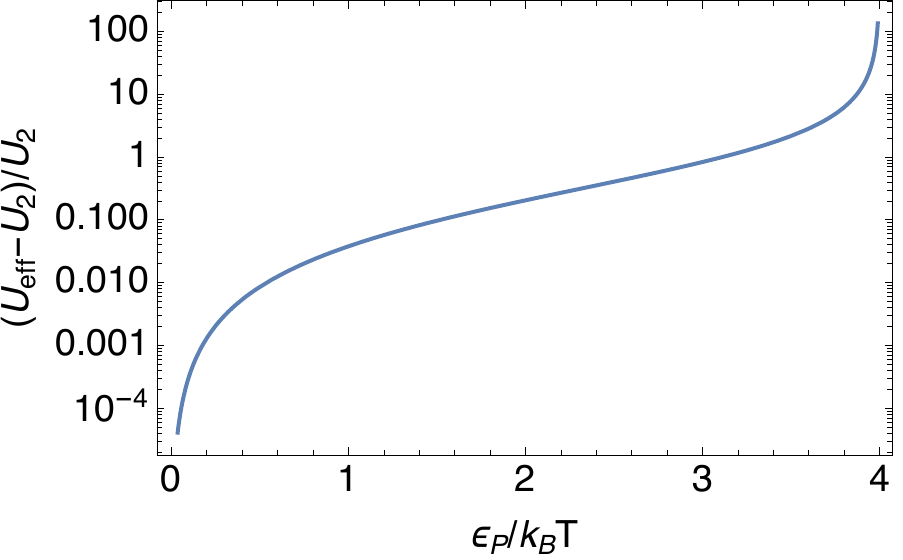}
\end{center}

\caption{
Fraction of thermal energy stored in the non-linearity, {\em i.e.} the ratio of the average of the sum of the non-linear terms of the potential to the average of the quadratic part $U_2$. 
}
\label{figUnl}
\end{figure}

The quantity $\langle U_{\rm eff} \rangle-U_2$ corresponds to the average of the sum of all the nonlinear terms of the potential, which could be interpreted as the amount of thermal energy stored in the nonlinearity.
One finds numerically that 
$\langle U_{\rm eff} \rangle-U_2 \approx  0.0169 {\tilde \epsilon_P} + 0.001 {\tilde \epsilon_P} ^2$
in the region $0 <  {\tilde \epsilon_P}  \le 4$. 
This quantity is  finite at ${\tilde \epsilon_P} =4$ where $U_2$ vanishes. 
Thus, approaching this value the sum of the contribution of the non-linear terms becomes dominant
with respect to the contribution of the quadratic term. 
A plot of $(\langle U_{\rm eff} \rangle-U_2)/U_2$ as a function of 
$\tilde \epsilon_P$ is shown in Fig.~\ref{figUnl}.
$(\langle U_{\rm eff} \rangle-U_2)/U_2 \approx 0.033 {\tilde \epsilon_P}^2$ for $\tilde \epsilon_P\rightarrow 0$ and 
$(\langle U_{\rm eff} \rangle-U_2)/U_2 \approx 1.34/(4-\tilde \epsilon_P)$ for $\tilde \epsilon_P\rightarrow 4$, the bistability threshold. 

\subsection{Nonlinear Duffing response in presence of thermal fluctuations}
\label{Duffing}

In this subsection we consider the response of the non-linear oscillator to an external drive. 
The results of this section are used to extract the value of the coupling $g$ from the measurement of the shift of the resonance frequency as a function of the driven vibration amplitude in Fig.~4b of the main text and to perform the fit of the response spectrum.
The measurements are performed as close as possible to the symmetric point, for which 
the potential is symmetric in $\tilde x$. 
We will thus focus on this symmetric point, limiting the expansion to the quartic term.
This corresponds to the standard Duffing oscillator in presence of thermal fluctuations. 
We find that the typical thermal amplitude of fluctuations of the oscillator are sufficiently large to
induce a sizable change in the resonance frequency. 
It is thus crucial to include these fluctuations that modify quantitatively the nonlinear
response to an external drive. 

We will follow standard methods to describe the system \cite{dykman_theory_1979,dykman_fluctuations_1980}. 
For clarity and uniformity of notation, we 
derive the main equations that lead to a Fokker-Planck description 
(see \refe{FokkerPlanck} in the following) of the slow degrees of freedoms: the two quadratures.
We begin by writing a Langevin equation for the displacement ${\tilde x}$:
\beq
	\ddot {\tilde x} =- \gamma \dot {\tilde x}   - \omega_0^2 {\tilde x} - \gamma_D {\tilde x}^3+ f_D \cos(\omega t) + f_N(t)
	,
\eeq
where $\gamma_D$ is the non-linear Duffing 
coefficient, and $f_D$  and $f_N(t)$ are the driving and thermal forces divided by the mass.
We assume
\beq
	\langle f_N(t) f_N(t') \rangle = 2 D \delta(t-t')
\eeq
with $D=k_B T \gamma/m$.
We now introduce the complex (quadrature) variable $z(t)$ as follows:
\beqa
	{\tilde x}(t)&=&  z(t) e^{i \omega t} + z(t)^* e^{-i \omega t} \label{Eqxt} \\
	\dot {\tilde x}(t)  	&=& i\omega \left[ z(t) e^{i \omega t} - z(t)^* e^{-i \omega t}\right] \label{Eqdxt}
	.
\eeqa
We can now perform the derivative of the above two equations:
\beqa
	\dot {\tilde x}(t)		&=&  i\omega \left[ z(t) e^{i \omega t} - z(t)^* e^{-i \omega t}\right]
	\label{Eqxt2}
	\\
	\ddot {\tilde x}(t)  	&=& 2 i\omega \dot z(t) e^{i \omega t} +( i\omega)^2 \left[ z(t) e^{i \omega t} +  z(t)^* e^{-i \omega t}\right]
	\nonumber \\
	&&
	\label{Eqdxt2}
\eeqa
where we have used the condition 
$
	\dot z(t) e^{i \omega t} + \dot z(t)^* e^{-i \omega t}  =0 
$ 
implicit in the definition of $z$.
Substituting  the equations for ${\tilde x}$, $\dot {\tilde x}$, and $\ddot {\tilde x}$ in the equation of motion,
multipling it by $e^{-i\omega t}$, and averaging it over a period with the assumption that $z$ evolves slowly gives
\beqa
	2i\omega \dot z(t) &=& 
	 \omega^2  z(t)  
	-i\gamma \omega  z(t) 
	-\omega_0^2 z(t)
	-3\gamma_D |z|^2 z(t) 
	\nonumber \\
	&&
	+{f_D \over 2} 
	+\langle f_N(t) e^{-i \omega t} \rangle
	.
\eeqa
We now introduce the time variable $\tau= \gamma t/2$ and scale $z$ as
$q(\tau)= \sqrt{3\gamma_D/\omega \gamma} z(2 \tau/\gamma)$.
This gives
 \beq
	\dot q(\tau)= 
	 -i \Omega  q(\tau)  
	-  q(\tau) 
	+i |q|^2 q(\tau) 
	-iF_0 -i F_N(\tau),
\eeq
 where we approximated $\omega^2-\omega_0^2\approx 2\omega(\omega-\omega_0)$ and defined
 \beqa
 	\Omega &=& {(\omega-\omega_0)\over \gamma/2},
 	\qquad
 	F_0 =  {\sqrt{3 \gamma_D}  f_D\over 2 \left(\omega \gamma\right)^{3/2}}	,
	\\
	F_N(t) & =& 
	\left\langle
	{\sqrt{3 \gamma_D} f_N(t) e^{-i\omega t}\over  \left(\omega \gamma\right)^{3/2}}	
	\right\rangle .
\eeqa

Neglecting the fluctuations, the stationary solution reads
\beq
	q_0 = {F_0/(|q_0|^2-\Omega+i)}.
\eeq
This defines the usual Duffing response. 
In particular one can express the dimensionless resonance frequency as a function of the amplitude:
\beq
    \Omega=|q_0|^2\pm \sqrt{F_0^2/|q_0|^2-1}
    \label{Omegaq}
    .
\eeq
The first term defines what is called the 'back-bone' of the resonance. 
This corresponds roughly to the dependence of the  maximum of the amplitude oscillation 
on the driving frequency when measuring the spectra for different drives. 
It depends quadratically on the dimensionless oscillation amplitude. 
When the thermal fluctuations are negligible 
this dependence can be used to extract the value of the 
non-linear Duffing coefficient from the data.
We discuss in the following the validity of this relation in presence of large thermal fluctuations.

We can now introduce $q=u+iv$, with $u$ and $v$ real. We have then
\beq
	\dot u  = g_u(u,v)+ \xi_u(\tau) ,   \qquad
	\dot v =  g_v(u,v)+ \xi_v(\tau) ,
\eeq
with 
\beq
	\xi_u+i\xi_v = - 
	i
	{\sqrt{3 \gamma_D} \left\langle f_N(t) e^{i\omega t}\right\rangle /\left(\omega \gamma\right)^{3/2}}	
	 .
\eeq
and 
\beq
	g_u =-u-(u^2+v^2-\Omega)v 
	,
	\qquad
	g_v=-v+(u^2+v^2-\Omega)u -F_0	
	.
\eeq
The correlation functions of the $\xi$-fields can be approximated by
$	\langle \xi_u(\tau) \xi_v(\tau') \rangle =0 $,
$
	\langle \xi_u(\tau) \xi_u(\tau') \rangle 
	= 
	\langle \xi_v(\tau) \xi_v(\tau') \rangle = 2 {\cal D} \delta (\tau-\tau')
$
where  
\beq
	{\cal D}  = {3 \gamma_D D \over 4 \omega^3 \gamma^2}={3 \gamma_D k_B T \over 4 m \omega^3 \gamma}
	\label{Ddef}
	.
\eeq
Finally from the two Langevin equations for $u$ and $v$ we can derive a Fokker-Planck 
equation for the probability $W(u,v)$:
\beq
	{\cal D} (\partial_u^2+\partial_v^2) W -\partial_u(g_u W)- \partial_v(g_v W) = \partial_t W
	.
	\label{FokkerPlanck}
\eeq
\rem{
In the absence of fluctuations it is easy to write the form of the response:
\beq
	\Omega= |q_0|^2\pm \sqrt{F_0^2\over |q_0|^2-1}
\eeq
it depends on a single parameter, $F_0$. One finds that a bistability appears for 
$F_0> 1.3$. For $F_0\ll1$ a standard Lorentian is found. 
}
The Fokker-Planck  \refe{FokkerPlanck} can be solved numerically for a given set of 
parameters to obtain the function 
\beq
	q_0(F_0,{\cal D}, \Omega)=u_0+iv_0
	= \int du dv \, W(u,v) (u+iv)
	.
	\label{defQuad}
\eeq
This gives the average of the two quadratures in  dimensionless units. 

\begin{figure}[h]
\begin{center}
 \includegraphics[width=8cm]{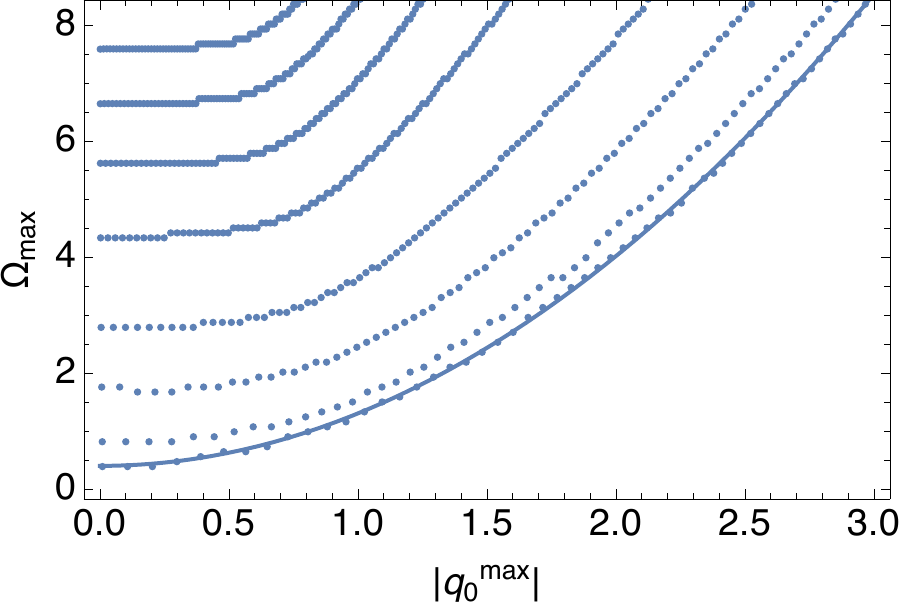}
 \end{center}
\caption{
The Figure shows $\Omega_{\rm max}$ as a function 
of $|q_0^{\rm max}|$ for ${\cal D}=.1, .2, .5, 1., 2., 3., 4., 5.$, from the lower to the upper curve. 
The effect of the fluctuations is to shift the initial value to higher frequency and to deform the dependence on $|q_0^{\rm max}|$. 
In the case of ${\cal D}=.1$ we show a fit with a quadratic dependence that gives a coefficient of $0.9 |q_0^{\rm max}|^2$.
The small steps are due to the discretization of the frequency in the numerical calculation.
}
\label{figQmax}
\end{figure}

As a first application of this equation we calculate numerically the maximum of the response amplitude of the oscillator $|q_0^{\rm max}|$ and the value 
$\Omega_{\rm max}$ for which this maximum appears. 
When fluctuations are negligible, for ${\cal D}\rightarrow 0$,
from \refe{Omegaq} one has $\Omega_{\rm max}=|q_0^{\rm max}|^2$.
In Fig.~\ref{figQmax} we plot $\Omega_{\rm max}$ as a function 
of $|q_0^{\rm max}|$ for different values of ${\cal D}$.
For the smallest value of ${\cal D}=0.1$ a good parabolic behavior is observed. 
Increasing ${\cal D}$ the curves flatten and deviations from the simple quadratic behavior are visible. 
This shows that using the quadratic dependence of the back-bone to extract the Duffing coefficient gives a qualitatively reasonable result when ${\cal D}<1$. 
In order to have a more reliable estimate, in the next subsection we discuss a fitting procedure that exploits the form of the response spectrum as predicted by the Fokker-Planck approach.

\subsection{Procedure used to fit the nonlinear Duffing response}
\label{Fitting procedure}

\begin{figure*}[tb!]
	\begin{center}
	    \includegraphics[width=\linewidth]{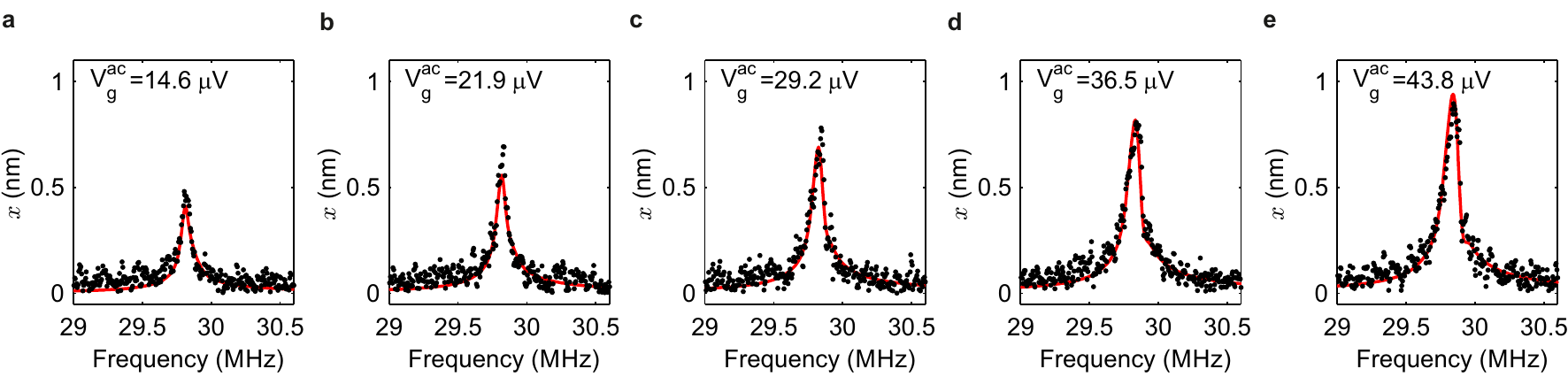}
		\caption{Driven nonlinear spectra spanning the linear-nonlinear crossover when increasing the drive at 6~K. We emphasize that a single set of parameters is used to fit the data of 10 different spectra; the fit is shown by the red line. 
		}
\label{fig:nonlinearspectra}
\end{center}
\end{figure*}

Using the results described in the previous sub-section, we now explain the procedure to determine the Duffing constant, and thus the ratio $\epsilon_P/k_BT$ and the coupling $g$, from driven nonlinear spectra when the thermal fluctuations are large. These data are shown in Figs.~4a,c in main text and Fig.~\ref{fig:nonlinearspectra}. 
We obtain the Duffing constant by collectively fitting the whole set of measured spectra spanning the linear-nonlinear crossover when sweeping the drive intensity. 
The spectra are measured nearby the conductance peak, that is, almost at the symmetric point.
From the experimentally measured spectra, the two quadratures $\{U_{ni}, V_{ni}\}$ are extracted for $N_v$ different values of the driving gate voltage $V_n^{ac}$ and for  $N_w$ different values of the driving frequency $\omega_{ni}$
%
 %
One has thus a set of $2N_vN_w$ values with $N_v=10$ and $N_{n}=300$. 
Using  the expression of the nonlinear coefficient given by \refe{gammaDuffing} and the definition of ${\cal D}$ given by \refe{Ddef} we can write a dimensionfull expression of the 
quadratures $\{u_e,v_e\}$:
\beqa
u_e &=&   {1 \over \tilde \epsilon_P} \sqrt{k_B T \gamma \over m \omega_0^3} 
	u_0(F_0, {\tilde \epsilon_P}^2\omega_0/64\gamma, (2\omega-\omega_0)/\gamma)
	,\\
	v_e &=&   {1 \over \tilde \epsilon_P} \sqrt{k_B T \gamma \over m \omega_0^3} 
	v_0(F_0, {\tilde \epsilon_P}^2\omega_0/64\gamma, (2\omega-\omega_0)/\gamma).
\eeqa
Here $u_0(F_0,{\cal D},\Omega)$ and $v_0(F_0,{\cal D},\Omega)$ are the (average of the) dimensionless quadratures obtained from the solution of the Fokker-Planck equation as defined by 
\refe{defQuad}.
We then define the $\chi^2$ function:
\beqa
	\chi^2 &=&\sum_{n=1}^{N_v}\sum_{i=1}^{N_w} 
	\left[\left(u_e(\omega_{ni},F_0=F_v V_n^{ac})-U_{ni}\right)^2
	+
	\right. \nonumber \\
	&&
	\left.
	\left(v_e(\omega_{ni},F_0=F_v V_n^{ac})-V_{ni}\right)^2
	\right]
	.
\eeqa
The free parameters of the fit are $\{\omega_0, \gamma, \tilde\epsilon_P, F_v\}$.
The factor $F_v$ is the relation between the dimensionless driving force intensity and the experimental driving voltage. 
This only assumes that the driving force increases linearly with the amplitude of the injected ac drive.
The best fit for $V_\mathrm{g}^\mathrm{dc}=0.7572$ V gives the values
$\epsilon_P/k_B T=0.22$,
$\omega_0/2\pi=29.7696$ MHz, $\gamma/2\pi=13.229$ kHz, 
$F_v=2.10897\cdot 10^5$ V$^{-1}$.
The fit is shown in Fig.~\ref{fig:nonlinearspectra}. 

\begin{figure}[h]
\begin{center}
 \includegraphics[width=8cm]{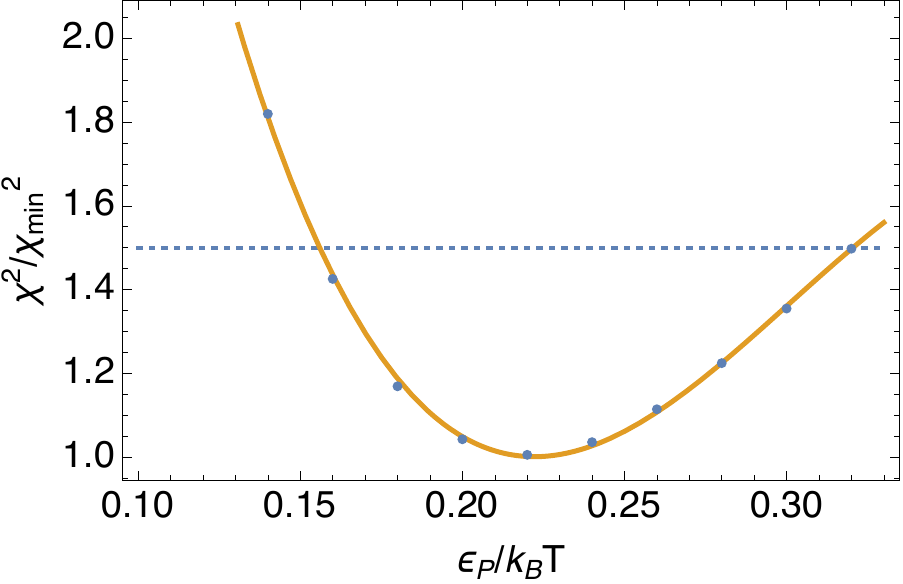}
 \end{center}
\caption{
Dependence of the $\chi^2$ on $\epsilon_P/k_B T$.
}
\label{figCHI2}
\end{figure}

In order to determine the error bar of the estimated value of $\epsilon_P/k_B T$, 
we find numerically the minimum of $\chi^2$ for a given value of $\epsilon_P/k_B T$ 
by tuning the other three parameters $\omega_0$, $\gamma$, and $F_v$ (see Fig.~\ref{figCHI2}).
We estimate this error by 
finding the range in $\epsilon_P/k_B T$ for which $\chi^2(\epsilon_P)<1.5 \chi_{\rm min}^2$,
where $\chi_{\rm min}$ is the minimum value of $\chi^2$.
The value of 1.5 is chosen so that the probability of observing a $\chi^2$ larger that this value is less than 1\%.
We find $ 0.15 <\epsilon_P/k_B T < 0.32$, which converts into a $\pm 120$ MHz error in the estimation of $g$.  
In addition to this error, one should include the imprecision in the calibration of the displacement, since the fit is very sensitive to the absolute value of the displacement. 
By performing the fit with different values of the displacement calibration,
we found that $\Delta \epsilon_P/k_B T \approx 0.15 \Delta A/A$,
where $\Delta A$ is the systematic error in the measurement of the displacement amplitude.
We estimate $\Delta A/A = 0.22$, which gives an additional $\pm 0.033$ to the error bar for $\epsilon_P/k_B T$ (and $\pm 97$~MHz error for $g/2\pi$). 
Overall, we get the coupling constant $g/2\pi=646 \pm 217$ MHz.
%
%

\section{Experimental Section}
\label{Experiments}
\subsection{Device production}

Carbon nanotubes are grown on high resistive silicon substrates with prefabricated platinum electrodes and trenches. The growth is done in the last step of the fabrication process to reduce surface contamination. Nanotubes are grown by the `fast heating' chemical vapour deposition (CVD) method, which consists in rapidly moving the sample from a position outside of the oven to the center of the oven, so that the temperature of the sample rapidly grows from room temperature to about 850\textdegree C. This enables us to grow nanotubes over shallow trenches~\cite{Bonis2018}. We remove the contamination molecules adsorbed on the nanotube surface during the transfer of the nanotube between the CVD oven and the cryostat, by applying a large current through the device under the ultra-high vacuum at the base temperature of the dilution cryostat~\cite{Yang2020}. In the three measured devices, the nanotube-gate separation is $150$~nm and the length of the suspended nanotube is between $1.2$~$\mu$m and $1.4$~$\mu$m.
\subsection{Electrical characterization}
\begin{figure*}[tb!]
	\begin{center}
		\includegraphics[width=0.9\linewidth]{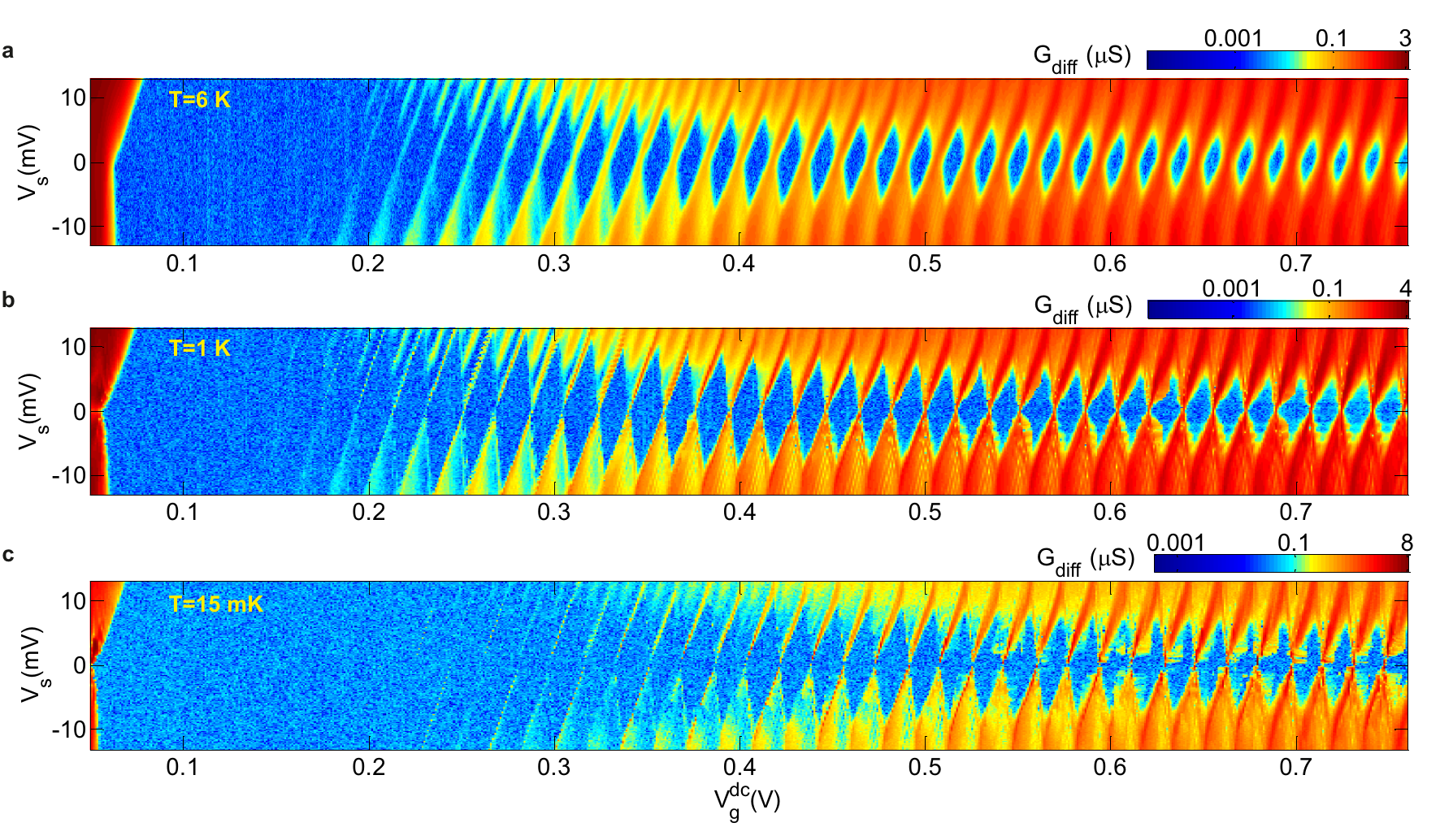}
		\caption{Differential conductance of device~I measured as a function of the source-drain voltage $V_\mathrm{s}$ and the gate voltage $V_\mathrm{g}^\mathrm{dc}$ at different temperatures. The temperature of the cryostat is 6~K, 1~K, and 15~mK.}
\label{diamonds}
\end{center}
\end{figure*}

We select 
ultra-clean, small-gap semiconducting nanotubes. 
Figures \ref{diamonds}a-c show charge stability diagram measurements at 6~K, 1~K, and the base temperature of the cryostat. The nanotube regions in contact with the source and drain electrodes are p-doped~\cite{Laird2015}. For large positive gate voltages, p-n junctions are formed along the nanotube near the metal electrodes, forming a quantum dot along the suspended nanotube. For gate voltage values below  $0.05$~V, the suspended nanotube region is p-doped and the p-n junctions disappear, resulting in a higher conductance. The size of the Coulomb diamonds decreases as the number of electrons in the nanotube quantum dot increases. 
The charging energy $E_\mathrm{c}$ varies approximately from $8.5$~meV to $6.5$~meV in the gate voltage range discussed in the main text, while the level spacing $\Delta E$ changes from $0.97$~meV to $0.73$~meV. All the data shown in the main text and supplementary materials are in the regime $ k_\mathrm{B}T< \Delta E, E_\mathrm{c}$. 
The short separation between the nanotube and the gate electrode enables us to achieve a large capacitive coupling between the nanotube island and the gate electrode $C_\mathrm{g} \gg C_\mathrm{s},C_\mathrm{d}$, where $C_\mathrm{s}$ and $C_\mathrm{d}$ are the capacitances between the nanotube island and the source and drain electrode, respectively. The diamonds in the charge stability diagram measurements become distorted when lowering temperature due to the mechanical self-oscillations of the suspended nanotube generated at finite source bias voltages~\cite{Steele2009,Urgell2020,Wen2020}.

\subsection{Temperature calibration}

\begin{figure}[h]
	\begin{center}
	    \includegraphics[width=\linewidth]{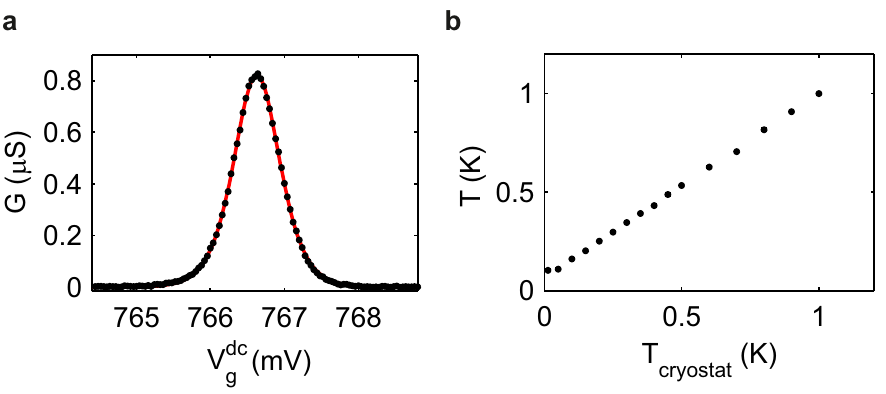}
		\caption{Temperature calibration. (a) Gate voltage dependence of the conductance of device I at $T=1$~K. The red solid line is the fit to the data using Eq.~\ref{GVg1K}. (b) The electron temperature of the device measured as a function of the cryostat temperature.}
\label{Gpeak}
\end{center}
\end{figure}

The temperature calibration in quantum dot devices operated in the incoherent single-electron tunneling regime ($\hbar\Gamma_\mathrm{e}<k_\mathrm{B}T< \Delta E, E_\mathrm{c}$) is achieved by measuring the electrical conductance peak (Fig.~\ref{Gpeak}a) where $\Gamma_\mathrm{e}$ is the electron coupling rate and $T$ is the temperature. The electron temperature is obtained from the gate voltage $V_\mathrm{g}^\mathrm{dc}$ width of the conductance peak using the standard incoherent single-electron tunnelling description [see \refe{Gtunn} in Sec.~\ref{IncTrans}]:
\begin{equation}
  {G=\dfrac {G_0}{\cosh^2 \Big[ \alpha (V_\mathrm{g}^\mathrm{dc}-V_\mathrm{P})/2k_\mathrm{B}T \Big] } } .
  \label{GVg1K}
\end{equation}
%
Here $G_0$ is the $T$-dependent peak conductance, $\alpha$ is the lever arm, and $V_\mathrm{P}$ is the gate voltage of the conductance peak.
We checked with numerical calculations of the Fokker-Planck equation that the modification of the width of the conductance peak by the electromechanical coupling is negligible over the measured temperature range. Figure~\ref{Gpeak}b shows that the electron temperature is linear with the cryostat temperature except at low temperature where it saturates at about 100~mK. 
We cannot estimate the temperature of the mechanical vibration fluctuations by measuring their spectrum as a function of temperature, since the low mechanical $Q$-factor due to the electron tunnelling in the SET regime impedes us to resolve the resonance of thermal vibrations. In Ref.~\cite{Urgell2020} we measured the vibration fluctuation temperature of a high-$Q$ nanotube device as a function of cryostat temperature using the same cryostat and the same cabling, filters, and amplifier; we observed that the vibration temperature is linear with the cryostat temperature down to a saturation temperature that is similar to the electron saturation temperature in Fig.~\ref{Gpeak}b. This indicates that the vibration temperature and the electron temperature are similar.

\subsection{Detection of mechanical vibrations and estimation of the effective mass}

\begin{figure}[h]
	\begin{center}
	    \includegraphics[width=\linewidth]{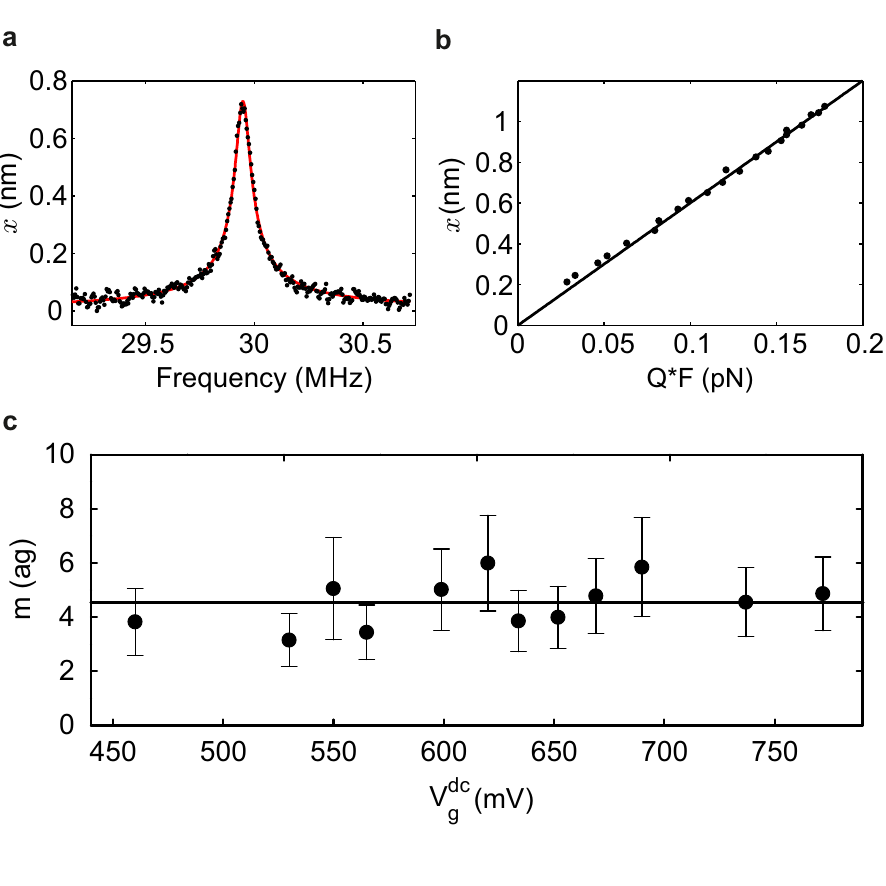}
		\caption{Determination of the effective mass. (a) Spectral response of the displacement amplitude of the driven vibrations at $T=6$~K. We chose the gate voltage close to the half maximum of the conductance peak, where the transconduction is largest. The red solid line is the fit of the data to a Lorenzian peak. (b) Force-displacement response curve at the mechanical resonance frequency. (c) Effective modal mass measured at different gate voltages. The black solid line indicates the average effective mass of $4.5$~ag.}
\label{masscalibration}
\end{center}
\end{figure}

\begin{figure*}[t!]
	\begin{center}
	    \includegraphics[width=0.9\linewidth]{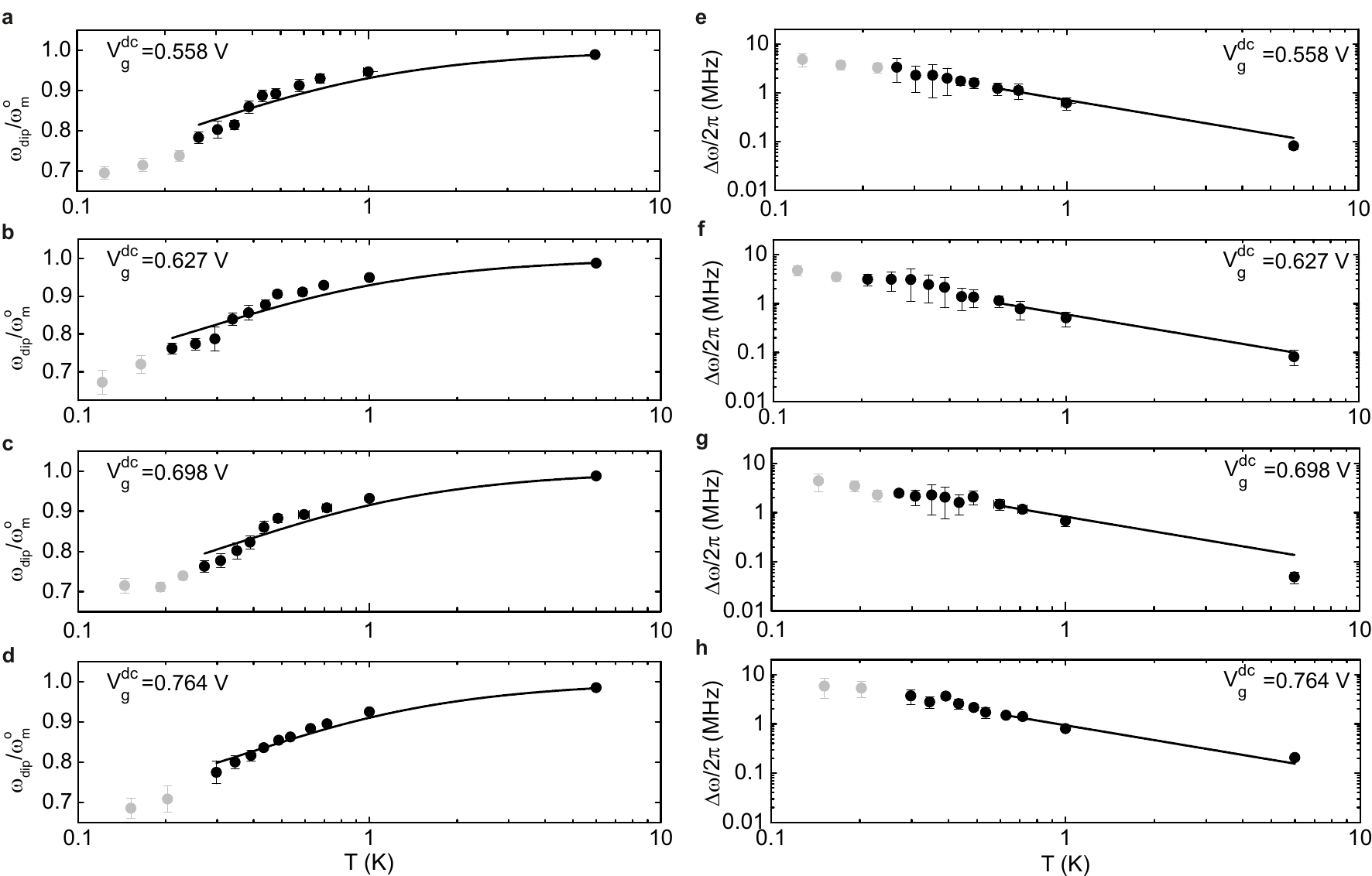}
		\caption{Determination of the electromechanical coupling and the electron tunnel rate. (a)-(d) Temperature dependence of the mechanical resonance frequency $\omega_\mathrm{dip}$ for different conductance peaks. We define $\omega_\mathrm{dip}$ as the lowest resonance frequency when sweeping the gate voltage over a conductance peak (Fig.~2b of main text). The gate voltage of the resonance frequency dip matches the gate voltage of the conductance peak. The black solid lines indicate the SET-based predictions. 
		(e)-(h) Temperature dependence of the mechanical resonance width for different conductance peaks. The black solid lines indicate the SET-based predictions in the high temperature limit. }
\label{fig:frequencysuppressionlinewidth}
\end{center}
\end{figure*}

We use a new two-source mixing method to measure the spectral mechanical response of driven vibrations in the linear and the nonlinear regimes. This method enables us to extract the spectral mechanical response by eliminating the inherent contribution of pure electrical origin in electrical mixing measurements~\cite{Samanta2022}. We detect the vibrations by capacitively driving them with an oscillating voltage with amplitude $V_\mathrm{g}^\mathrm{ac}$ and frequency $\omega$, applying the oscillating voltage with amplitude $V_\mathrm{g}^\mathrm{ac}$ with a slightly detuned frequency $\omega+\delta\omega$ on the source electrode, and measuring the current at frequency $\delta\omega$ from the drain electrode. To improve the sensitivity of the current detection, we connect the drain electrode to a $RLC$ resonator placed nearby the device and a HEMT amplifier at the 4~K stage of the dilution cryostat~\cite{Bonis2018}. The $RLC$ resonator enables us to measure the current at a comparatively high frequency $\delta\omega\simeq 1.2$~MHz where the $1/f$ noise is significantly reduced. Without the inductance $L$ of the $RLC$ resonator, the frequency $\delta\omega$ has to be set to a lower frequency, typically $1-10$~kHz, within the bandwidth imposed by the resistance of the sample and the capacitance of the electrical cables that connect the device to the measurement instruments. 
To obtain the spectral mechanical response of driven vibrations, we separate the signal of the mechanical vibrations from the signal of pure electrical origin inherent to the mixing method by properly tuning the phase of the measured current~\cite{Samanta2022}. This is important since the measured current is otherwise a non-trivial combination of the vibration displacement and the electrical contribution. The pure electrical contribution is used as a resource to calibrate the signal of the vibrations into units of meters. Figure~\ref{masscalibration}a shows the spectral response of driven vibrations, which can be well described by a Lorentzian peak.

The effective mass of the measured mechanical eigenmode can be reliably determined, since the driven vibration amplitude can be calibrated with the two-source mixing method described above and since the capacitive force in quantum dot electromechanical devices can be accurately quantified. The mass $m$ is quantified from the slope of the force-displacement ($F$-$x$) response at the mechanical resonance frequency in Fig.~\ref{masscalibration}b using $x=(Q/m\omega_\mathrm{m}^2)F$ where the quality factor $Q$ is estimated from the spectral response in Fig.~\ref{masscalibration}a and the capacitive force is given by
\begin{align}
\label{eq:force}
&F=\beta C^{\prime}_\mathrm{g}V_\mathrm{g}^\mathrm{dc}V_\mathrm{g}^\mathrm{ac},\\
\label{eq:BetaForce}
&\beta=1-\dfrac{C_\mathrm{g}}{C_\mathrm{\Sigma}} +f(1-f)\dfrac{C_\mathrm{g}}{C_\mathrm{\Sigma}} \dfrac{e^2/C_\mathrm{\Sigma}}{k_\mathrm{B}T},
\end{align}
in the incoherent SET regime~\cite{Samanta2022}. The dot-gate capacitance $C_\mathrm{g}$, the total capacitance $C_\mathrm{\Sigma}$ of the dot, and the average charge occupation number $f$ (with value between 0 and 1) are all quantified by standard electron transport measurements. The spatial derivative of the dot-gate capacitance $C_\mathrm{g}'$ is determined from $C_\mathrm{g}$ and the dot-gate separation $d$ using the expression of the capacitance between a cylinder and a plate that leads to $C_\mathrm{g}'=C_\mathrm{g}/d\ln{(2d/r)}$. Figure~\ref{masscalibration}c shows the effective mass measured at twelve different conductance peaks. The average effective mass is $m=4.5\pm1.5$~ag. The uncertainty in the mass determination comes from the mass fluctuations in Fig.~\ref{masscalibration}c and the uncertainty in the estimation of the dot-gate separation. We estimate the nanotube radius $r=1.5$~nm from the measured mass and the suspended nanotube length determined by scanning electron microscopy. This value is consistent with the radii of the nanotubes produced with our chemical vapour deposition growth.

\subsection{Electromechanical coupling and electron tunnel rate }

\begin{figure*}[tb!]
	\begin{center}
	    \includegraphics[width=0.9\linewidth]{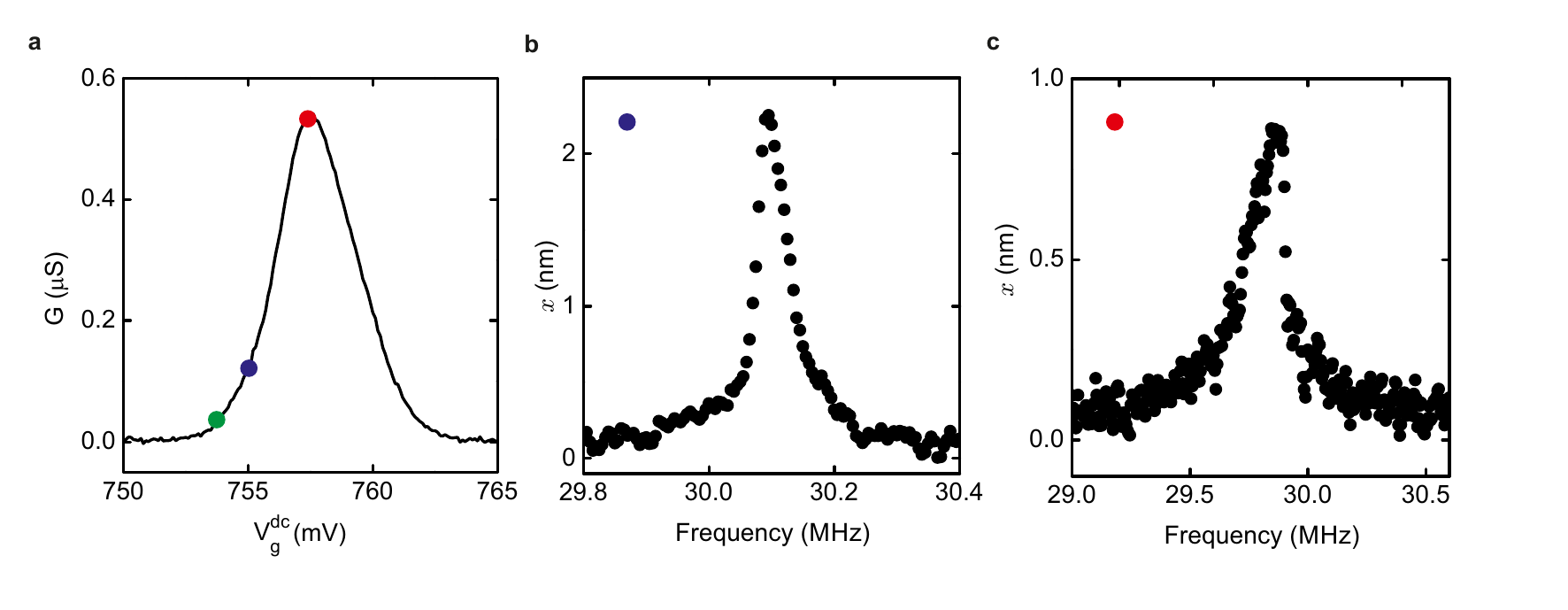}
		\caption{Spring hardening and softening in the nonlinear spectral response of mechanical vibrations in Device~I at 6~K. (a) Gate voltage dependence of the conductance. (b) Nonlinear response showing spring softening when the system is set at the base of the conductance peak (blue point). (c) Nonlinear response showing spring hardening when the system is set at the top of the conductance peak (red point).}
\label{fig:posnegDuffing}
\end{center}
\end{figure*}

Figures~\ref{fig:frequencysuppressionlinewidth}a-h show the temperature dependence of both the resonance frequency and the resonance width of driven vibrations measured at the conductance peaks for different gate voltages. The fitting of these data to the predictions of the theory enable us to determine the coupling $g$ and the total electron tunnel rate $\Gamma_\mathrm{e}$ for these different conductance peaks. The values of $g$ and $\Gamma_\mathrm{e}$ are shown in Figs.~3a,c of the main text. In the fitting we only select the black data points in Figs.~\ref{fig:frequencysuppressionlinewidth}a-d with resonance frequency ratios $\omega_\mathrm{dip}/\omega_\mathrm{m}^\mathrm{o}$ between $0.75$ and $1$, since it is the range of values expected by the predictions discussed in Sec.~\ref{SpectrumSoft}. %
The grey data points correspond to data at lower temperature where a double-well potential is expected to emerge, but further work in needed to characterize this regime. The coupling of vibrations and SET also results in dissipation. The mechanical resonance width in the high temperature limit ($k_\mathrm{B}T\gg\hbar g^2/{\omega_\mathrm{m}}$) is given by 
\begin{equation}
  {\Delta\omega= \Gamma_0 + \dfrac {1}{2} \dfrac {\hbar g^2}{k_\mathrm{B}T} \dfrac {\omega_\mathrm{m}^\mathrm{o}}{\Gamma_\mathrm{e}}}   
  \label{eq:dissipationSET}
\end{equation}
where $\Gamma_0$ is the damping due to other dissipation mechanisms, see Eq.~\ref{SETdissipation}. We fit the measured resonance width in the high temperature limit with Eq.~\ref{eq:dissipationSET} in Fig.~\ref{fig:frequencysuppressionlinewidth}e-h. The electron tunnel rates $\Gamma_\mathrm{e}$ obtained from the fits are shown in Fig.~3c of the main text. 

\subsection{Nonlinear spectral response of mechanical vibrations}

We show here that the nonlinear Duffing constant measured at 6~K changes sign when sweeping the gate voltage through the conductance peak, in agreement with the predictions of the theory in Sec.~\ref{NonLinearCoef}. This enables us to rule out other possible origins for the nonlinearity, such as the geometrical nonlinearity~\cite{Bachtold2022}.
We observe both a softening and hardening spring behaviour of the oscillator over a narrow range in gate voltage, see Figs.~\ref{fig:posnegDuffing}a-c. Figure~\ref{fig:posnegDuffing}c shows the spectral response of the spring hardening when the system is set at the conduction peak as indicated by the red dot in Fig.~\ref{fig:posnegDuffing}a. By contrast, Fig.~\ref{fig:posnegDuffing}b shows the spring softening effect at the base of the conductance peak as highlighted by the blue dot in Fig.~\ref{fig:posnegDuffing}a. The change of the nonlinear Duffing sign is consistent with the predictions of the theory describing the coupling of mechanical vibrations and SET in the incoherent regime. Indeed, Eq.~\ref{gammaDuffingfirst} indicates that the Duffing constant is positive at the conductance peak when the average charge occupation fraction $f=1/2$, while it becomes negative at the base of the peak when $f$ is sufficiently close to zero.

\subsection{Responsivity of mechanical vibrations}

\begin{figure}[H]
	\begin{center}
	    \includegraphics[width=\linewidth]{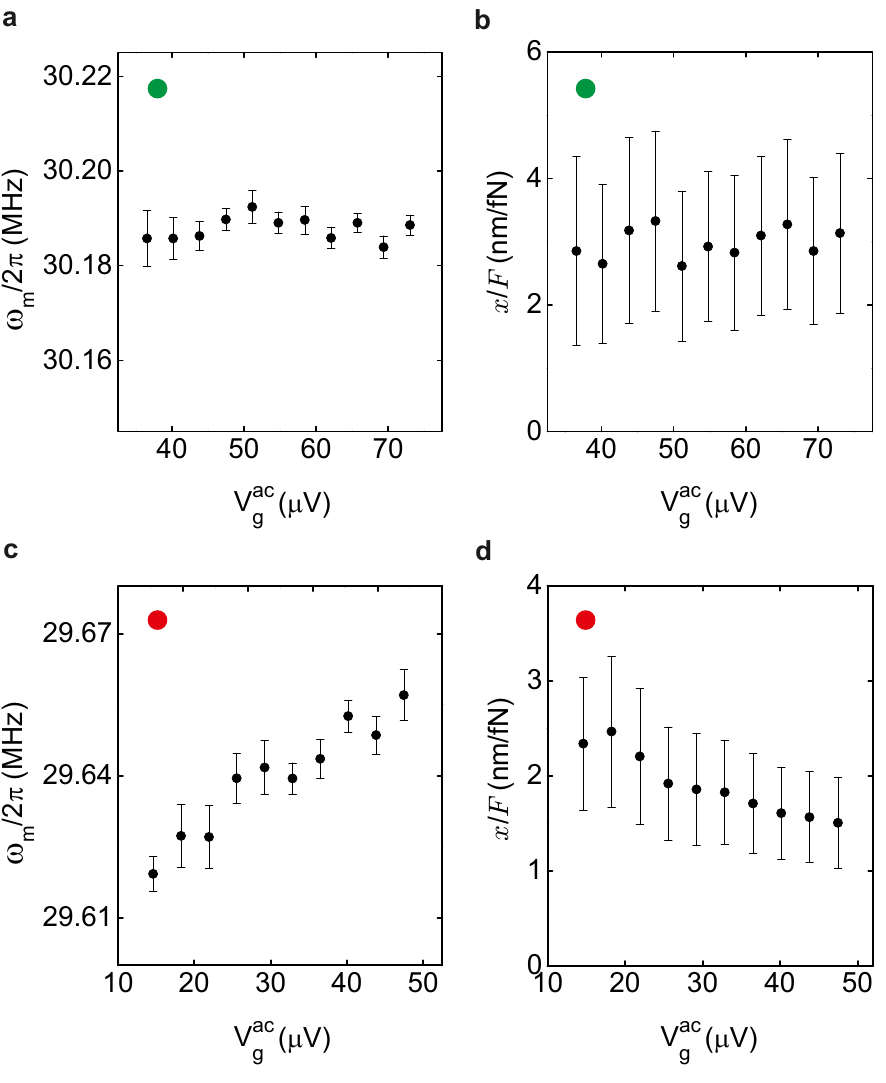}
		\caption{Responsivity of mechanical vibrations at 6~K. (a,b) Resonance frequency and responsivity of the vibrations as a function of the driving voltage with the system set at the very base of the conductance peak in Fig.~\ref{fig:posnegDuffing}a (green dot). (c,d) Same as (a,b) but with the system set on the top of the conduction peak (red dot).}
\label{fig:responsivity}
\end{center}
\end{figure}

We show that the observed reduction of the responsivity at large drive is related to the thermal switching between coexisting stable states in driven nonlinear oscillators. Figures~\ref{fig:responsivity}a,b show the drive dependence of both the mechanical resonance frequency and the responsivity of the mechanical vibrations when the system is set at the very base of the conductance peak, see green dot in Fig.~\ref{fig:posnegDuffing}a.

We do not observe any shift in resonance frequency, indicating that the Duffing constant is becoming small. This is expected from Eq.~\ref{gammaDuffingfirst} when $f\simeq 0$, that is, when the effect of the coupling between vibrations and SET is suppressed. We do not observe any change of the responsivity either, showing that nonlinear damping plays a negligible effect~\cite{Lifshitz2008}.

By contrast, Fig.~\ref{fig:responsivity}d shows that the responsivity gets lower when increasing the drive in the case where the system is set at the top of the conductance peak, see red dot in Fig.~\ref{fig:posnegDuffing}a. This reduction is well reproduced by the SET-based predictions (Fig.~4d in main text), which relates this behaviour to switching between coexisting stable states in driven nonlinear oscillators, and not to nonlinear damping.

\begin{figure}[tb!]
	\begin{center}
	    \includegraphics[width=\linewidth]{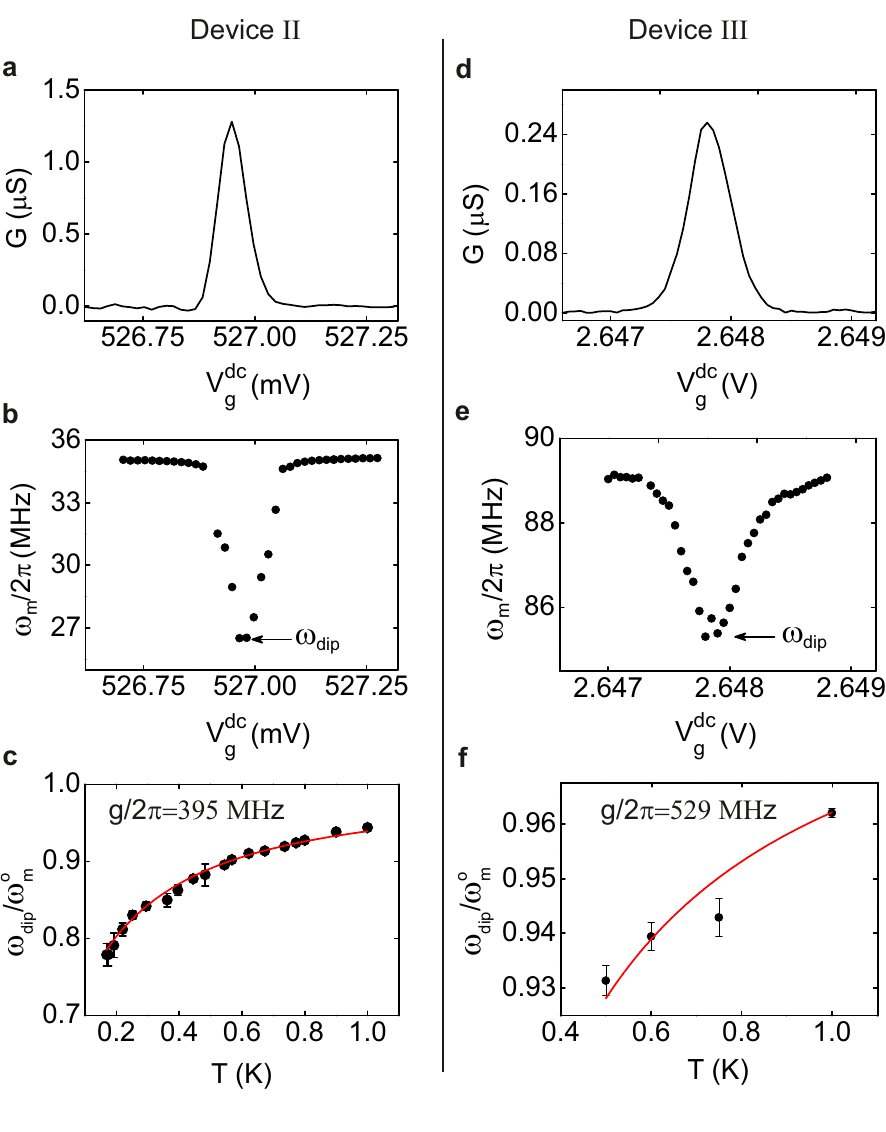}
		\caption{Strong anharmonicity and ultra-strong coupling regime in two other devices. (a,b) Conductance and mechanical resonance frequency as a function of gate voltage at 170~mK for Device~II. (c) Suppression of the resonance frequency as a function of temperature at the conduction peak. The solid red line is the SET-base prediction. (d,e) Conductance and mechanical resonance frequency as a function of gate voltage at 6~K for Device~III. (f) Suppression of the resonance frequency as a function of temperature at the conduction peak.}
\label{fig:otherdevices}
\end{center}
\end{figure}

\subsection{Strong anharmonicity in two other devices}
%

We demonstrate strong anharmonicity and ultra-strong coupling regime in two other devices. Figure~\ref{fig:otherdevices}a shows a conduction peak of Device II. The charging energy, the level spacing, and the electron tunnel rate are $E_\mathrm{c}=14$~meV, $\Delta E = 2$~meV, and $\hbar \Gamma_\mathrm{e}=2$~$\mu$eV, respectively, showing that SET is in the incoherent regime ($\hbar\Gamma_\mathrm{e} <k_\mathrm{B}T< \Delta E, E_\mathrm{c}$). The high lever arm $\alpha=0.83$ arises from the short separation between the nanotube and the gate electrode. 
Figure~\ref{fig:otherdevices}b shows a dip with a large suppression of the mechanical resonance frequency when sweeping the gate voltage through the conductance peak; the bare resonance frequency is $\omega_\mathrm{m}^\mathrm{o}/2\pi=35.1$~MHz. Figure~\ref{fig:otherdevices}c shows the temperature dependence of the mechanical resonance frequency at the dip. The ratio $\omega_\mathrm{dip}/\omega_\mathrm{m}^\mathrm{o}$ approaches 0.75 at low temperature, indicating that the potential of the vibrations becomes strongly anharmonic. The red solid line is the fit of the data to the SET-based predictions, enabling us to quantify $g/2\pi=384$~MHz. This value is similar to the value $g/2\pi=395$~MHz obtained with independently measured parameters and using $g=e(C_\mathrm{g}'/C_\mathrm{\Sigma})V_\mathrm{g}^\mathrm{dc}/\sqrt{2m\hbar\omega_\mathrm{m}^\mathrm{o}}$. These data indicate that the device is in the ultrastrong coupling ($g > \omega_\mathrm{m}^\mathrm{o}$) and the adiabatic regime ($\Gamma_\mathrm{e} > \omega_\mathrm{m}^\mathrm{o}$), which satisfy
the conditions to realize strong vibration anharmonicity.
Figures~\ref{fig:otherdevices}d-f show the data of Device~III. We obtain $g/2\pi=529$~MHz and $\omega_\mathrm{m}^\mathrm{o}/2\pi=89$~MHz, which shows that device is in the ultrastrong coupling regime. The measured suppression of the resonance frequency $\omega_\mathrm{dip}/\omega_\mathrm{m}^\mathrm{o}$ reaches 0.93 at 500~mK. The device could not be measured at lower temperature due to technical problems unrelated to the device itself. The anharmonicity is sizeable but not as large as that in Devices~I and II.

\subsection{Estimation of error bars in the figures of the main text}

In Fig. 2 of the main text,  the confidence interval error bars in panels (b) and (c) arise primarily from the standard deviation in $\omega_\mathrm{m}$ quantified from different driven spectral response measurements. In Fig. 3 of the main text, the confidence interval error bars in panels (a) (black dots) and (c) arise primarily from the uncertainty in the fit of the measured temperature dependence of $\omega_\mathrm{m}$ and $\Delta\omega$, respectively, to the predictions of the theory. The confidence interval in the estimation of $g$ shown in (a) (purple shaded area) mainly originates from the uncertainty in the measurement of the mass. In Fig. 4 of the main text, the confidence interval error bars in panels (b) and (c) arise from the uncertainty in the fitting of the spectral response and the determination of the dot-gate separation.

\subsection{Parameters of Device I discussed in the main text}

\begin{center}
\begin{tabularx}{0.46\textwidth} { 
  | >{\raggedright\arraybackslash}X 
  | >{\raggedleft\arraybackslash}X | }
 \hline
 \textbf{Parameters} & \textbf{Device I} \\
 \hline
 Suspended nanotube length (L)  & $1.2$ $\mu$m  \\
  \hline
 Nanotube radius (r)  & $1.5$ nm \\
 \hline
 Effective mechanical mode mass (m) & $4.5$ ag  \\
  \hline
 Bare resonance frequency ($\omega_\mathrm{m}^\mathrm{o}/2\pi$) & $28.3$-$30.3$ MHz  \\
\hline
 Nanotube-gate separation (d)  & $150$ nm  \\
 \hline
 Zero point fluctuation ($x_\mathrm{zp}$) & $7.9$ pm  \\
 \hline
 Nanotube-gate capacitance ($C_g$)  & $9.7$ aF  \\
\hline
 Lever arm ($\alpha$)  & $0.4-0.5$  \\
\hline
 Charging energy ($E_c$)  & $8.5-6.5$ meV  \\
\hline
 Level spacing ($\Delta E$)  & $0.97-0.73$ meV \\
\hline
 Work function difference between nanotube and gate & $120$ mV \\
\hline

\end{tabularx}
\end{center}
\vspace{5 cm}
